\documentclass[%
 aip,
 amsmath,amssymb,
 reprint,%
]{revtex4-1}

\usepackage{graphicx}% Include figure files
\usepackage{dcolumn}% Align table columns on decimal point
\usepackage{bm}% bold math
\usepackage[utf8]{inputenc}
\usepackage[T1]{fontenc}
\usepackage{mathptmx}
\usepackage{etoolbox}

\usepackage[usenames,dvipsnames, pdftex]{xcolor}

\makeatletter
\def\@email#1#2{%
 \endgroup
 \patchcmd{\titleblock@produce}
  {\frontmatter@RRAPformat}
  {\frontmatter@RRAPformat{\produce@RRAP{*#1\href{mailto:#2}{#2}}}\frontmatter@RRAPformat}
  {}{}
}%
\makeatother
\begin{document}

\preprint{AIP/123-QED}

\title{Subtle Nuances between Quantum and Classical regimes}

\author{Karin Wittmann Wilsmann}%
\affiliation{Instituto de F\'{i}sica, Universidade do Rio Grande do Sul, RS  91501-970, Brazil}
\email{karin.wittmann@ufrgs.br}%
\author{Erick R. Castro}%
\affiliation{Centro Brasileiro de Pesquisas F\'{i}sicas/MCTI, RJ 22290-180, Brazil}
%\email{erickc@cbpf.br}
\author{Itzhak Roditi}%
\affiliation{Centro Brasileiro de Pesquisas F\'{i}sicas/MCTI, RJ 22290-180, Brazil}
%\email{roditi@cbpf.br}
\author{Angela Foerster}%
\affiliation{Instituto de F\'{i}sica, Universidade do Rio Grande do Sul/ UFRGS, RS 91501-970, Brazil}
 %\email{angela@if.ufrgs.br}
\author{Jorge G. Hirsch}%
\affiliation{Instituto de Ciencias Nucleares, Universidad Nacional Aut\'{o}noma de M\'{e}xico, Cd. Mx. 04510, Mexico}
 \email{hirsch@nucleares.unam.mx}

%%%%%%%%%%%%%%%%%%%%%%%%%%%%%%%%%%%%%%%%%%%%%%%%%%%%%

\date{\today}

\begin{abstract}
This study explores the semiclassical limit of an integrable-chaotic bosonic many-body quantum system, providing nuanced insights into its behavior. 
We examine classical-quantum correspondences across different interaction regimes of bosons in a triple-well potential, ranging from the integrable to the self-trapping regime, and including the chaotic one. 
The close resemblance between the phase-space mean projections of classical trajectories and those of Husimi distributions evokes the Principle of Uniform Semiclassical Condensation (PUSC)
of Wigner functions of eigenstates. Notably, the resulting figures also exhibit patterns reminiscent of Jason Gallas's "shrimp" shapes.
\end{abstract}

\maketitle

\begin{quotation}
Understanding the connection between quantum (microscopic) and classical (macroscopic) behaviors in particle systems is a key question in theoretical physics and is essential for predicting system evolution and designing experiments. In this work, we explore these relationships for bosons confined in a triple-well potential, where quantum technologies enable precise control.
Through phase-space projections, we observe remarkably similar trajectories in both frameworks, revealing a strong quantum-classical correspondence across regular and chaotic dynamics. This correspondence reinforces the Principle of Uniform Semiclassical Condensation of Husimi functions, which posits that the wavefunction of a quantum system in phase space approximates classical trajectories in the semiclassical limit. Notably, the resulting patterns evoke the 'shrimp' shapes described by J.A.C. Gallas.
\end{quotation}

%-------------------------------------------%
%-------------------------------------------%

\section{\label{sec:intro}INTRODUCTION}
Natant decapod crustaceans, commonly known as shrimps (and prawns), are ubiquitous species inhabiting oceans and shallow seas worldwide, including the frigid waters of the Antarctic. These crustaceans play a crucial role in marine ecosystems, serving as predators, scavengers, and prey. Their diverse feeding habits and ecological roles contribute significantly to nutrient cycling and energy flow within marine environments.
In the field of dynamical systems, the term "shrimps" was introduced by J.A.C. Gallas~\cite{GALLAS1993,GALLAS1994,HuntGallas1999,BonattoGallas2005,BonattoGallas2008}. This terminology arises from the visual similarity between certain bifurcation diagrams and the shape of these marine animals. Bifurcation diagrams, which map the points at which a system's behavior changes, often display complex, branching structures that resemble the segmented bodies and antennae of shrimps. Interestingly, these dynamical systems' "shrimps" are ubiquitous structures, appearing across a wide range of nonlinear systems. Their presence highlights the intricate and often unexpected patterns that can emerge in the study of dynamical systems, underscoring the universality of such phenomena.

In tribute to J.A.C. Gallas, here we unveil patterns that evoke his iconic "shrimp" shapes in quantum maps for a Bose-Hubbard model, with a clear correspondence to classical behavior.

Our approach focuses on a generalized three-site Bose-Hubbard model with open boundary conditions and long-range interactions~\cite{Lahaye2010},

\begin{align}
{\mathcal H}& = \frac{U_0}{2} \sum_{i=1}^{3}  N_i (N_i-1) + \sum_{i=1}^{3} \sum_{j=1; j\neq i}^{3} \frac{U_{ij}}{2} N_i N_j \nonumber \\ 
&\qquad - J_1(a_1^\dagger a_2 +  a_2^\dagger a_1) - J_3(a_2^\dagger a_3 + a_3^\dagger a_2 ).
\label{bhm}
\end{align}
where $a_i^\dagger$, $a_i$,  $i=1,2,3$, are the canonical creation and annihilation operators, representing the three bosonic degrees of freedom in the model, $N_i=a_i^\dagger a_i$ the number operators of the well $i$.
The coupling $J_i,\, i=1,3$  
denotes the tunneling between neighboring wells, and $U_0$ and $U_{ij}=U_{ji}$, $i\neq j$, set the on-site and long-range dipole–dipole interactions, respectively.
The Hamiltonian has two independent conserved quantities: the energy and the total number of particles $N$, with $N=N_1+N_2+N_3$.
An experimental feasibility of this system was detailed in Ref. \cite{Lahaye2010} (see also Ref. \cite{Wittmann2018}).

For a given set of parameters, the model exhibits a third conserved quantity~\cite{Ymai2017}, equating to the number of degrees of freedom. 
In this configuration the model is integrable, presenting a bipartite structure~\cite{Tonel2005, Links2006}. 
Systems with a two-mode algebra exhibit three distinct regimes of interaction~\cite{Lahaye2010}:
 {\it Rabi}:  $U \ll JN^{-1}$; 
 {\it Josephson}: $JN^{-1} \ll U \ll JN$; and
{\it Fock}: $JN \ll U$.
In the Rabi and Josephson regimes, the behavior is semiclassical, 
while the Fock regime corresponds to a pendulum in a strongly quantum regime  \cite{Lahaye2010}.
The semiclassical integrability properties of the 3-well model have been extensively studied in the Josephson regime \cite{Wittmann2018, Tonel2020, wittmann2023}, as well as in a 4-well model \cite{grun2022interfer, grun2022protocol}, both of which belong to a family of integrable multimode systems\cite{Ymai2017}.

Conversely, integrable quantum models become even more intriguing when they can be controllably driven to chaos~\cite{Marissa2021}.
Studies have shown that this integrable system can be driven to the chaotic limit for a finite number of particles \(N\), although its eigenstates do not achieve maximum ergodicity, earning it the label "Preface to many-body quantum chaos"~\cite{Castro21,Wittmann2022,CastroPRA2024}. 
The chaotic behavior is achieved by breaking the symmetry of the system through a tilt between potential wells. Additionally, the system must be in the transition between the Rabi and Josephson interaction regimes.
Chaos has also been explored in other three-well models~\cite{Nemoto2000,Franzosi2001,Mossmann2006,Graefe2006,Hiller2006,Liu2007,Hiller2009,Kollath2010,Viscondi2011,March2018,Bera2019,Rautenberg2020,Ray2020,Nakerst2021,NH2022,Bhattacharyya2023,Zhou24}.

In this study, we explore the correspondence between the classical and quantum behaviors of a 3-well bosonic system as it transitions from integrability to chaos and then to a fully localized regime. A main finding is that the phase-space mean projections of classical trajectories and those of Husimi distributions evoke the Principle of Uniform Semiclassical Condensation (PUSC) of Wigner functions of eigenstates\cite{Robnik2019,Robnik2020}. 
The PUSC states that Wigner or Husimi quasiprobabilty distributions of eigenstates condense uniformly on a classical invariant component in the classical phase space, when the Heisenberg time is larger than all relevant classical transport time scales~\cite{Robnik2019,Robnik2020}. 
We employ measurements designed to visually capture these correspondences. Interestingly, the results bring to mind the distinct “shrimp” shapes highlighted by J.A.C. Gallas.

The article is organized as follows: In Sec. \ref{sec:system}, we introduce the integrable model along with its symmetry-breaking term, and we also present its classical counterpart. In Section \ref{sec:inttochaos}, we compare quantum eigenvector projections (condensations) with classical trajectories of the system as integrability is broken. Sec. \ref{sec:chaos} extends this quantum-classical analysis to subspaces of the system in the chaotic regime. 
n Sec. \ref{sec:special}, we present some trajectories exhibiting quantum-classical correspondence that drew our attention due to their resemblance to Gallas's "shrimps."
Finally, Sec. \ref{discussion} provides a discussion of the results, addresses key questions, and outlines new perspectives for future research.

%-------------------------------------------%
%-------------------------------------------%

\section{\label{sec:system}SYSTEM DESCRIPTION}

\subsection{\label{sec:qmodel}Quantum model}	
\begin{figure}
    \centering
 \includegraphics[width=1\linewidth]{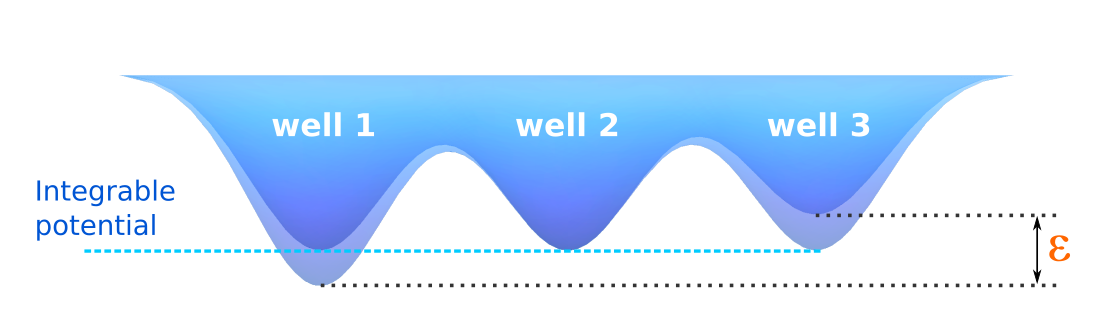}
    \caption{Diagram illustrating the effect of the external potential $\epsilon$ on wells 1 and 3, used to break integrability.}
    \label{fig:1}
\end{figure}

For some parameter values ~\cite{Ymai2017}
the Hamiltonian \eqref{bhm} becomes integrable and can be expressed in the reduced form \cite{Wittmann2018}
	\begin{align}
	\label{QH}
	\hat{H} =& \frac{U}{N}\left(\hat{N}_1-\hat{N}_2+\hat{N}_3\right)^2 + \epsilon\left(\hat{N}_3-\hat{N}_1\right) \nonumber \\ &+\frac{J}{\sqrt{2}}\left(\hat{a}_1^\dagger \hat{a}_2 + \hat{a}_2^\dagger \hat{a}_1\right)+\frac{J}{\sqrt{2}}\left(\hat{a}_2^\dagger \hat{a}_3 + \hat{a}_3^\dagger \hat{a}_2\right),
	\end{align}
where a breaking term $\epsilon\left(\hat{N}_3-\hat{N}_1\right)$ has been added and a constant term is ignored. 
Here, the parameter $U$
% $U=(U_{12}-U_0)/4$ 
represents the coupling constant for inter-site and intra-site interactions, and an isotropic tunneling $J_1=J_3=J/\sqrt{2}$ was adopted~\cite{Wittmann2018,Tonel2020}.
The parameter \(\epsilon\) represents the amplitude of an external potential that generates a tilt between wells 1 and 3. The Hamiltonian is integrable for \(\epsilon = 0\). In this case, in addition to the energy and the total number of particles \(N\), there is a third independent conserved quantity, expressed through the operator \(Q = J_1^2 N_3 + J_3^2 N_1 - J_1 J_3(a_1^\dagger a_3 + a_3^\dagger a_1)\), which can be interpreted as a two-well subsystem involving wells 1 and 3. When \(\epsilon \neq 0\), the Hamiltonian becomes non-integrable, reaching the maximum degree of chaoticity for \(\epsilon \sim J, U\)~\cite{Wittmann2022}. Hereafter, units are chosen such that $\hbar=1$. A scheme of the action of the external potential on potential wells is shown in Fig. \ref{fig:1}. 

The Hamiltonian matrix has a dimension of \(D = \frac{(N+2)!}{2!N!}\). Its eigenstates are represented in the Fock basis, defined as \(|N_1, N_2, N_3 \rangle \). The eigenvalues of \(\hat{H}\) are denoted as \( E_{m} \) and its eigenvectors by \( |m\rangle \).

The quantum analysis of the system is carried out using the Husimi function projected onto the Fock basis. The averaged Husimi function $\mathfrak{H}_m\left(N_1,N_3\right)$ of an eigenstate \( |m\rangle \), with eigenenergy \( E_m \), projected onto the Fock component \((N_1, N_3)\), is given by~\cite{CastroPRA2024}
\begin{equation}\label{eq:HusFock}
\mathfrak{H}_m\left(N_1,N_3\right)=\frac{1}{\mathcal{N}}\sum_{n = m - \frac{\mathcal{N}}{2}}^{m + \frac{\mathcal{N}}{2}}\vert \langle N_1, N-N_1-N_3, N_3 \vert n \rangle \vert^2,
\end{equation}
where \(\mathcal{N}\) is a sufficiently large number of states with energies close to a reference energy \(E_m\).
In our case, we use \(\mathcal{N}=200\).

The averaged Husimi function, projected onto the Fock basis, serves as a valuable tool, as it not only offers computational advantages but also strengthens the connection between the classical and quantum regimes.

For simplicity, we will omit the hat notation for the quantum operators \(\hat{N_i}\) from now on.

%-------------------------------------------%
%-------------------------------------------%

\subsection{\label{sec:clmodel}Classical model}
The classical Hamiltonian can be derived from \(\hat{H}\) \eqref{QH} using \(\mathcal{H}_{\text{cl}} = \frac{\langle \alpha |\hat{H} |\alpha \rangle}{N}\), where \(|\alpha\rangle\) represents the coherent states \(|\alpha\rangle = |\alpha_1, \alpha_2, \alpha_3 \rangle\), with \(\alpha_k = \sqrt{N_k} \exp(i\phi_k)\), where \(\sqrt{N_k}\) and \(\phi_k\) represent the amplitude and phase of the coherent state for the mode \(k = 1, 2, 3\), respectively. This yields:
	\begin{align}
 \bar{\mathcal{H}}_{\text{cl}} &= \frac{U}{N}\left(N_1-N_2+N_3\right)^2 + \epsilon\left(N_3-N_1\right) \nonumber \\ + &J \sqrt{2} \left[\sqrt{N_1N_2}\cos(\phi_1-\phi_2)+\sqrt{N_2N_3}\cos(\phi_2-\phi_3)\right].\nonumber 
	\end{align}

By defining \(\rho_k = \sqrt{N_k/N}\), with the condition \(\rho_1^2 + \rho_2^2 + \rho_3^2 = 1\), we have $\rho_2=\sqrt{1 - \rho_1^2 - \rho_3^2}$, and the classical Hamiltonian can be reduced to the simplified form
\begin{align}\label{eq:hclass1}
{\mathcal{H}}_{\text{cl}} =& U\left( 2(\rho_1^2+\rho_3^2)-1\right)^2 + \epsilon\left(\rho_3^2-\rho_1^2\right) \nonumber \\ 
+&J \sqrt{2}  \sqrt{1 - \rho_1^2 -\rho_3^2} \left[\rho_1 \cos(\phi_{12})+ \rho_3 \cos(\phi_{23})\right]
\end{align} 
with $\phi_{ij}=\phi_i-\phi_j$. An initial condition of the classical system is defined by the variables $P=(N_1/N, N_3/N, \phi_{12}, \phi_{32})$, which leaves the Hamiltonian ${\mathcal{H}}_{\text{cl}}$ with energy $E_{classic}={\mathcal{H}}_{\text{cl}}(N_1/N, N_3/N, \phi_{12}, \phi_{32})$.

For the dynamical evolution of these initial conditions, it is convenient to express Eq. \eqref{eq:hclass1} in terms of the conjugate coordinates \( Q_i = \frac{\alpha_i + \alpha_i^*}{\sqrt{2N}} \) and \( P_i = \frac{\alpha_i - \alpha_i^*}{i\sqrt{2N}} \), resulting in the form~\cite{CastroPRA2024}

\begin{align}
	{\mathcal{H}}_{\text{cl}} =\frac{\bar{\mathcal{H}}_{\text{cl}}}{N}=& \frac{U}{4} \left(Q_1^2+P_1^2-Q_2^2-P_2^2+Q_3^2+P_3^2\right)^2 \nonumber \\ 
	&+ \frac{\epsilon}{2}\left(Q_3^2+P_3^2-Q_1^2-P_1^2\right) \nonumber \\ 
	&+\frac{J}{ \sqrt{2}} \left[Q_1 Q_2 + P_1 P_2 + Q_2 Q_3 + P_2 P_3\right],\label{CH3}
\end{align} 
with the dynamics given by
\begin{equation}\label{eq:traj1}\left(\dot{Q}_i,\dot{P}_i\right)=\left(\frac{\partial \mathcal{H}_{\text{cl}}}{\partial P_i},-\frac{\partial 	\mathcal{H}_{\text{cl}}}{\partial Q_i}\right).\end{equation} 
The classical occupation coordinates \(N_i\) is recovered at each time using
\begin{equation}\label{eq:traj2}
    \frac{N_i(t)}{N} = \frac{Q_i^2(t) + P_i^2(t)}{2}.
\end{equation}

{\it Parameters:} The quantum-classical correspondence of the system is explored using the parameters \( U = 0.7 \) and \( J = 1 \) associated with the chaotic regime~\cite{Castro21,Wittmann2022,CastroPRA2024}. The chaotic regime is reached for $\epsilon_c=1.5$. Under these conditions, a critical classical energy $E_c \approx 0.075$ arises, corresponding to the unstable critical point $P_c \approx (0.081, 0.294, 0, \pi)$.

The correspondence between quantum and classical systems will be primarily visual, represented through Husimi projections (Eq. \eqref{eq:HusFock}) and classical trajectories (Eqs. \eqref{eq:traj1} and \eqref{eq:traj2}), both in the \(N_1\) and \(N_3\) coordinates.

%-------------------------------------------%
%-------------------------------------------%

\section{\label{sec:QCtraj}Quantum vs Classical trajectories}

\subsection{\label{sec:inttochaos}From integrability to chaos}
We begin by analyzing the behavior of the integrable system as the symmetry between wells 1 and 3 is broken by increasing the parameter \(\epsilon\). This analysis is first conducted using the Hamiltonian \eqref{QH} through the Husimi function \eqref{eq:HusFock}. Specifically, we consider the projection of the mean value of 200 eigenvectors with eigenvalues closest to the unstable classical critical energy corresponding to each value of \(\epsilon\)~\footnote{Through Eq. \eqref{eq:hclass1}, with \({\mathcal{H}}_{\text{cl}} = E_{\mathrm{classic}}\), we determine suitable initial conditions for classical trajectories with energy \(E_{\mathrm{classic}}\). A variant of Eq. \eqref{eq:hclass1}, incorporating the coordinate \(\rho_2\) and Lagrange multipliers, is employed to find the classical critical configurations and their corresponding energies~\cite{Castro21}.}.

\begin{figure}[h]
    \centering
     \includegraphics[width=12.5cm]{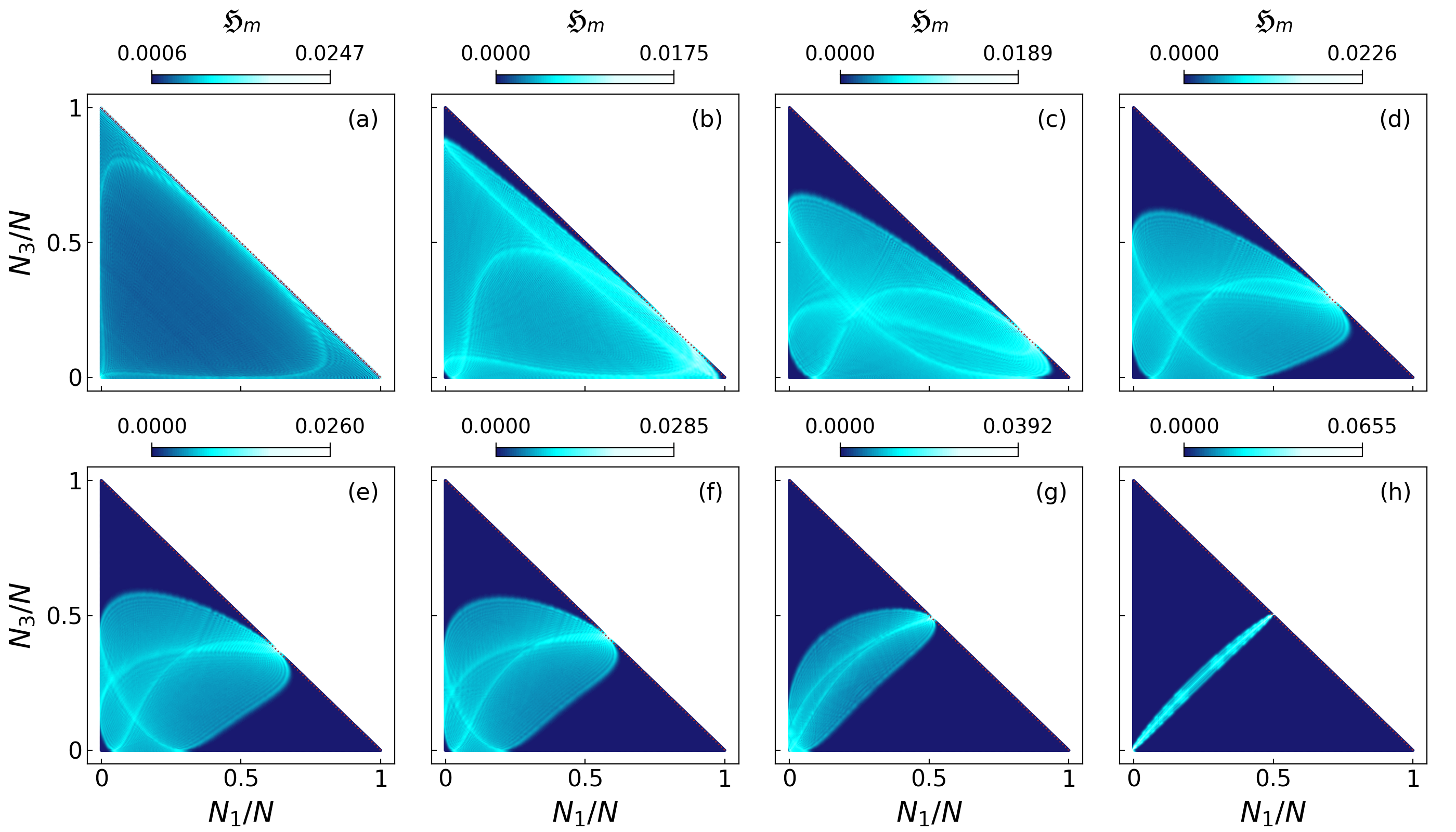}
    \caption{Representation of \(\hat{H}\) as it varies with \(\epsilon\) through the Husimi function \(\mathfrak{H}_{E}(N_1, N_3)\), projected onto the \(N_1\) vs \(N_3\) coordinates for \(N = 300\). Each panel shows the mean distribution of 200 eigenvectors with energies near the classical critical energy \(E_{\mathrm{classic}}\), which depends on \(\epsilon\). Panels: (a) \(\epsilon = 0\), (b) \(\epsilon = 0.5\), (c) \(\epsilon = 1\), (d) \(\epsilon = 1.5\), (e) \(\epsilon = 2\), (f) \(\epsilon = 2.5\), (g) \(\epsilon = 5\), (h) \(\epsilon = 30\). The symmetry of the integrable Hamiltonian in (a) is reflected between \(N_1\) and \(N_3\), gradually breaking up to the critical value \(\epsilon_c = 1.5\) (e), after which it tends to self-trap at \(N_1 = N_3\), as seen in (h).}
    \label{fig:Hevol}
\end{figure}

Figure \ref{fig:Hevol} illustrates how the Hamiltonian evolves with \(\epsilon\), starting from the integrable model at \(\epsilon = 0\) and progressing to \(\epsilon \gg U,J\). The symmetry of the integrable model, clearly visible in Fig. \ref{fig:Hevol}(a), gradually breaks as \(\epsilon\) increases, reaching the chaotic limit at \(\epsilon_c = 1.5\) in Fig. \ref{fig:Hevol}(d). Beyond this point, the trajectories tend back towards symmetry between wells 1 and 3, but with a self-trapped characteristic. Notably, in Fig. \ref{fig:Hevol}(a), the trajectory shows a diagonal alignment, tending towards \(N_2=N\). By Fig. \ref{fig:Hevol}(h), the trajectory aligns transversely, tending towards \(N_1=N_3\). 
This sequence highlights the transition from the integrable regime to a self-trapped regime, through the chaotic region~\cite{wittmann2023,CastroPRA2024}.

%------------------------------------------%

\onecolumngrid

\newpage

\twocolumngrid

\begin{figure}[t]
    \centering
          \includegraphics[width=12.5cm]{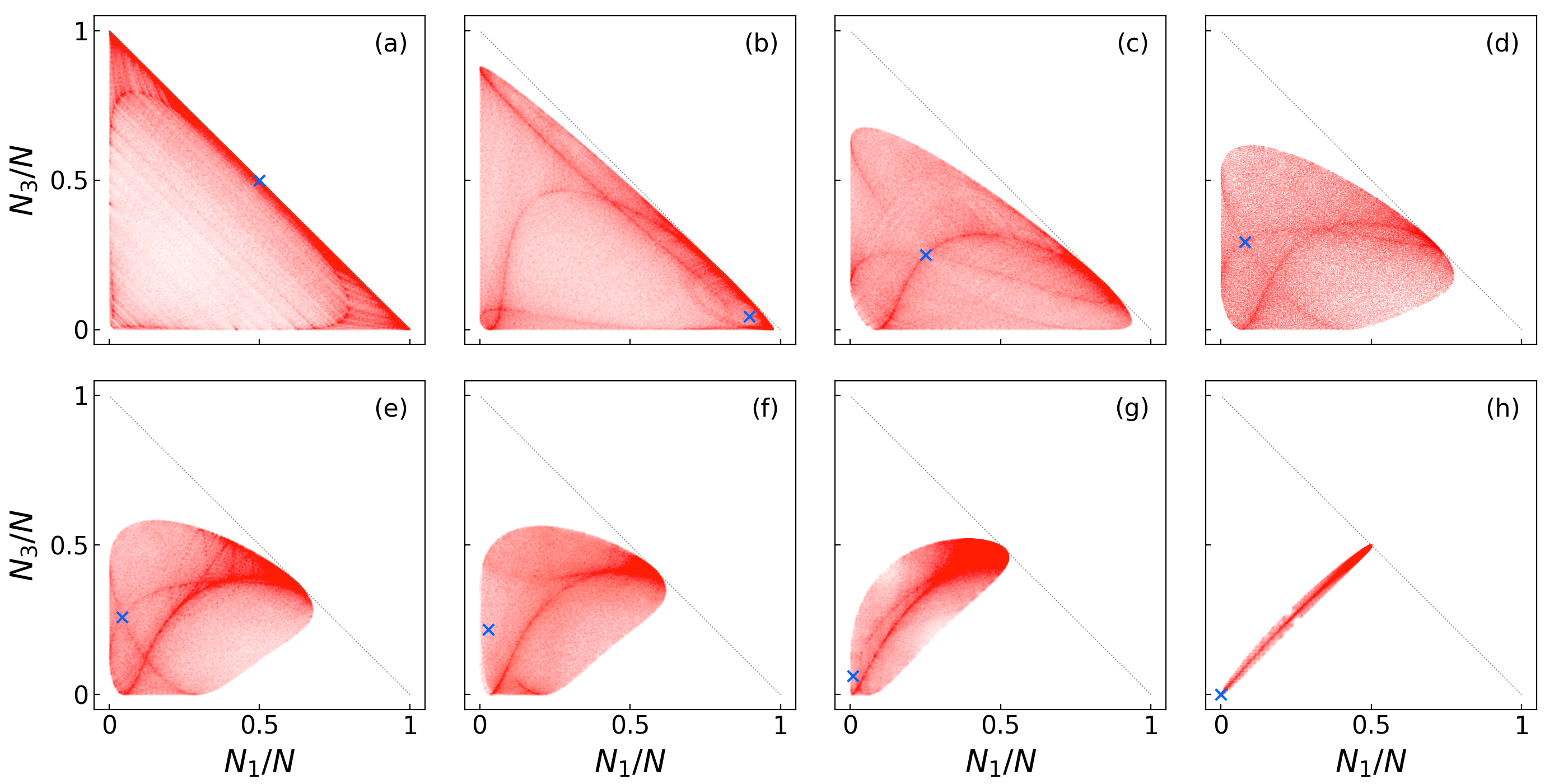}
    \caption{Representation of classical trajectories of \({\mathcal{H}}_{\text{cl}}\) as a function of \(\epsilon\) in coordinates \(N_1\) and \(N_3\), using the same parameters as Fig. \ref{fig:Hevol}, with \(\epsilon\) values ranging from (a) \(\epsilon = 0\), (b) \(\epsilon = 0.5\), (c) \(\epsilon = 1\), (d) \(\epsilon = 1.5\), (e) \(\epsilon = 2\), (f) \(\epsilon = 2.5\), (g) \(\epsilon = 5\), and (h) \(\epsilon = 30\). For each value of \(\epsilon\), the classical system exhibits a distinct critical energy, with a critical point marked with "x", and the initial conditions are chosen accordingly. Panels (a-c) and (g-h) show superimposed trajectories for various initial conditions at the critical energy, while panels (d-f) display single long-time evolved trajectories. The correspondence with the Husimi projections in Fig. \ref{fig:Hevol} is noticeable.}  
    % Representation of classical trajectories of \({\mathcal{H}}_{\text{cl}}\) as a function of \(\epsilon\), in coordinates \(N_1\) and \(N_3\), for the same parameters as in Fig. \ref{fig:Hevol}, with (a) \(\epsilon = 0\), (b) \(\epsilon = 0.5\), (c) \(\epsilon = 1\), (d) \(\epsilon = 1.5\), (e) \(\epsilon = 2\), (f) \(\epsilon = 2.5\), (g) \(\epsilon = 5\), and (h) \(\epsilon = 30\). 
    % For each value of \(\epsilon\), the classical system exhibits a distinct critical energy, with a critical point marked with "x",
    % and the initial conditions are chosen accordingly. 
    % Panels (a-c) and (g-h) display a superposition of multiple trajectories for different initial conditions, all corresponding to the respective critical energy. In contrast, panels (d-f) show trajectories for single initial conditions evolved over a long time, corresponding to their respective critical energies. The correspondence between these classical trajectories and the Husimi projections shown in Fig. \ref{fig:Hevol} is quite significant.}
    \label{fig:Classicevol}
\end{figure}

In Fig. \ref{fig:Classicevol}, we show the behavior of the semiclassical model for the same parameters considered in Fig. \ref{fig:Hevol}, through trajectories in phase space coordinates \((N_1, N_3)\).
Figures \ref{fig:Classicevol}(a-c) and \ref{fig:Classicevol}(g-h) show trajectories of the non-chaotic system, where the values of \(\epsilon\) are far from the chaotic parameter \(\epsilon_c = 1.5\). For each of these cases, it was necessary to consider the superposition of several trajectories, all with the same energy but different initial conditions \((N_1, N_3, \phi_{12}, \phi_{32})\), because individual trajectories tend to be localized.
Even so, the superposition of different trajectories partially reproduces the quantum average patterns of Fig. \ref{fig:Hevol}. Figures \ref{fig:Classicevol}(d-f) represent trajectories for \(\epsilon\) values closer to the chaotic parameter \(\epsilon_c\). Unlike the previous cases, these were generated with a single trajectory evolved over long periods and show a better correspondence with the quantum mean values shown in Fig. \ref{fig:Hevol}. In particular, the best correspondence occurs exactly for \(\epsilon_c\), Figs. \ref{fig:Hevol}(d) and  \ref{fig:Classicevol}(d). Here, any initial condition with the same energy generates the same trajectory, demonstrating the system's chaotic and ergodic properties.

%------------------------------------------%

\onecolumngrid

\twocolumngrid

%-------------------------------------------%
%-------------------------------------------%

\subsection{\label{sec:chaos}Through the chaos}

\begin{figure}[h!]
    \centering
   \includegraphics[width=0.9\linewidth]{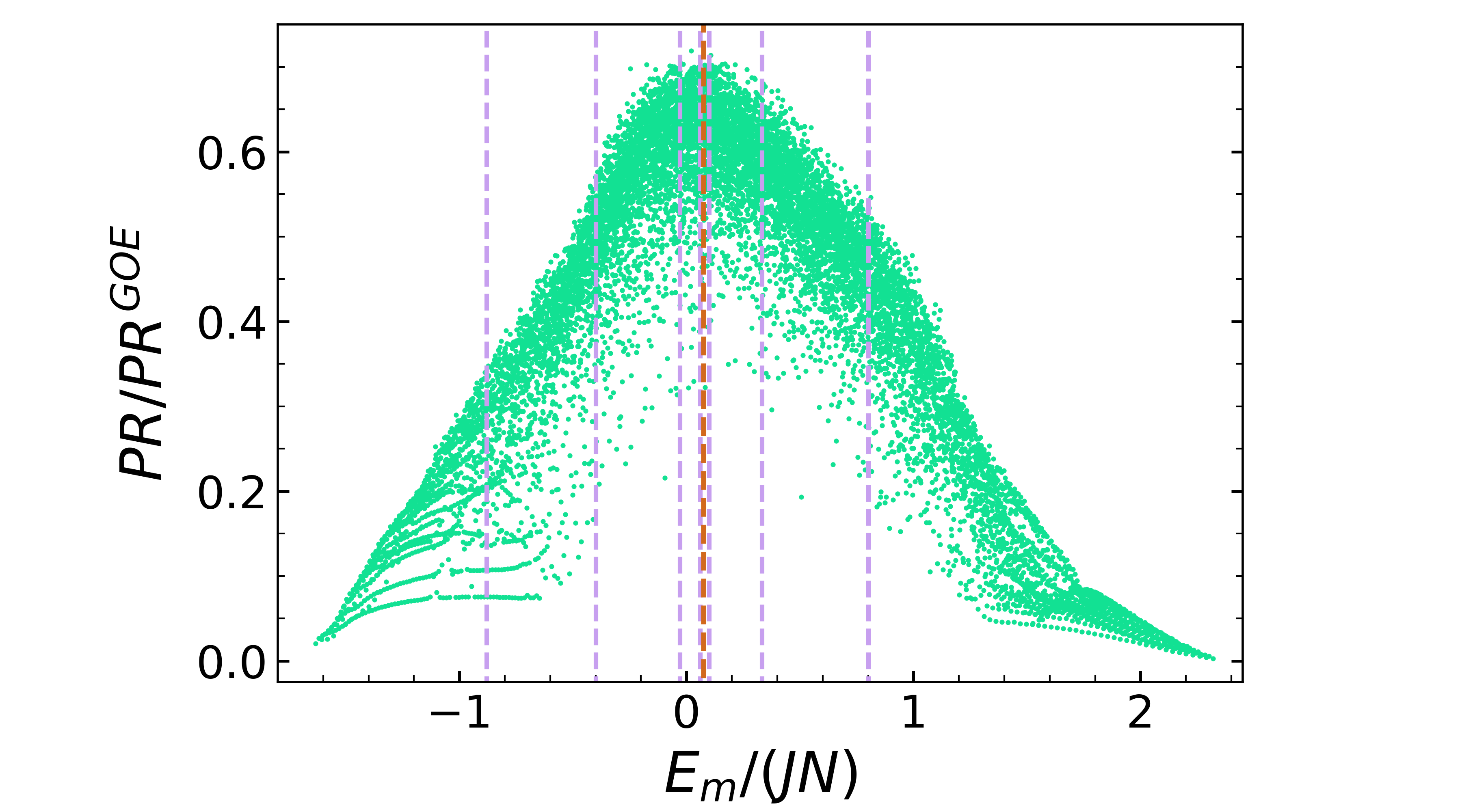}
    \caption{Participation ratio as a function of energy \(E_m\) for \(\hat{H}\), using the chaotic parameters \(U = 0.7\), \(J = 1\), and \(\epsilon_c = 1.5\) (same as in Figs. \ref{fig:Hevol}(d) and \ref{fig:Classicevol}(d)) for \(N = 150\). The dashed vertical lines mark the energies \(E_m \simeq -0.9, -0.4, -0.03, 0.06, 0.075, 0.1, 0.3\), and \(0.8\), which will serve as reference in Figs. \ref{fig:QHeps15} and \ref{fig:Classicevol2}. The brown dashed line highlights the critical energy \(E_{\mathrm{c}}\).}
    \label{fig:PR}
\end{figure}

In this section, we examine the system under chaotic conditions, where \(\epsilon_c = 1.5\), \(U = 0.7\), and \(J = 1\), as used in Figs. \ref{fig:Hevol}(d) and \ref{fig:Classicevol}(d).
For the quantum case, we analyze sets of eigenstates from different regions of the spectrum of the Hamiltonian \eqref{QH}. These regions are marked in Fig. \ref{fig:PR} by vertical lines overlaid on the distribution of the participation ratio of the system. The participation ratio (PR) is a well-known measure of delocalization \cite{Santos2016}, defined as:
\begin{equation}
PR^{m} \equiv \frac{1}{\sum_{n=1}^D|C^{m}_n|^4},
\label{Eq:PR}
\end{equation}
where \(C^{m}_n = \langle n | E_m \rangle\).
A system approaches chaos when the coefficients of its eigenstate components tend toward a homogeneous distribution, leading to an increase in the participation rate (PR). Eigenstates are considered to be fully delocalized when their distribution in Hilbert space is nearly uniform, with \(|C_n^m|^2\) fluctuating around \(1/D\). In contrast, integrable systems are characterized by regular, predictable behavior, and their energy eigenstates tend to be more localized in phase space. This leads to lower participation ratio values, indicating that fewer basis states contribute significantly to each eigenstate.

In Fig. \ref{fig:PR} the participation ratio of the eigenstates in the Fock basis is presented, scaled~\cite{Wittmann2022} by the expected value for a Gaussian Orthogonal Ensemble (GOE) random matrix, given by  \( \text{PR}^{GOE} \sim D/3 \)~\cite{mehta2004random}.

We can see that the participation ratio decreases at the spectrum edges, where stable critical points are located~\cite{Castro21}, while well-defined lines emerge, indicating localized states. Additionally, the region with the highest participation ratio is observed around the unstable critical energy, indicated by the brown vertical line in the figure.
The participation ratios of \(\hat{H}\) for the other values of \(\epsilon\) presented in the previous section are shown in App. \ref{app:PR}, where the classical critical energy for each case is also indicated.

\begin{figure}
    \centering
    \includegraphics[width=12.5cm]{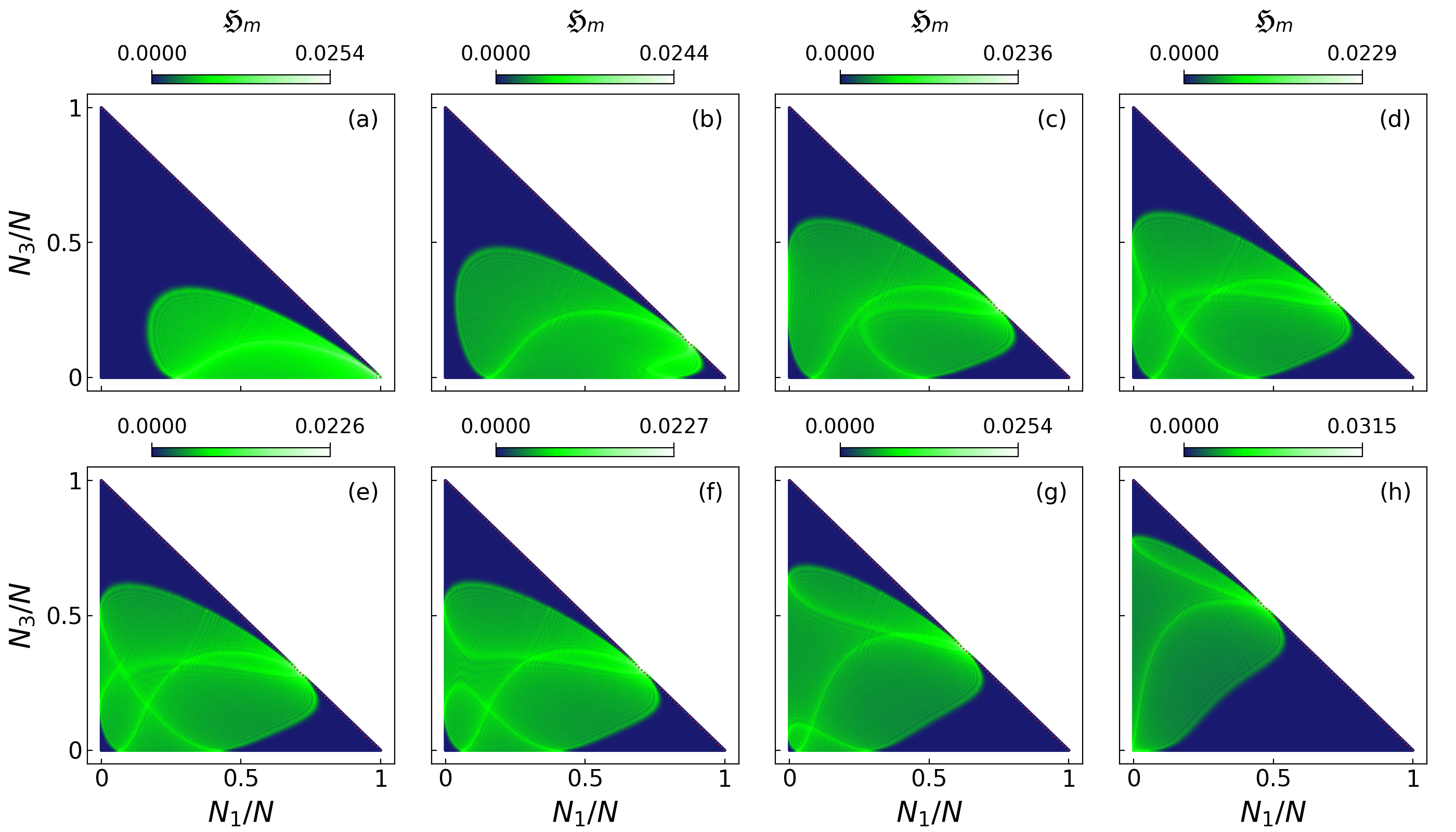}
    \caption{Representation of the quantum Hamiltonian \(\hat{H}\) for \(\epsilon_c = 1.5\) using the Husimi function \(\mathfrak{H}_{E}(N_1, N_3)\) across different ranges of eigenstates in the spectrum, for \(N = 300\). Each panel illustrates the distribution of the average value of 200 eigenvectors with eigenvalues near \(E_m \simeq -0.9\) (a),\; \(-0.4\) (b),\; \(-0.03\) (c),\; \(0.06\) (d), \(0.075\) (e),\; \(0.1\) (f),\;\(0.3\) (g)\; and \(0.8\) (h) (see also Fig. \ref{fig:PR}).}
    \label{fig:QHeps15}
\end{figure}

\begin{figure}[h!]
 \centering
    \includegraphics[width=12.5cm]{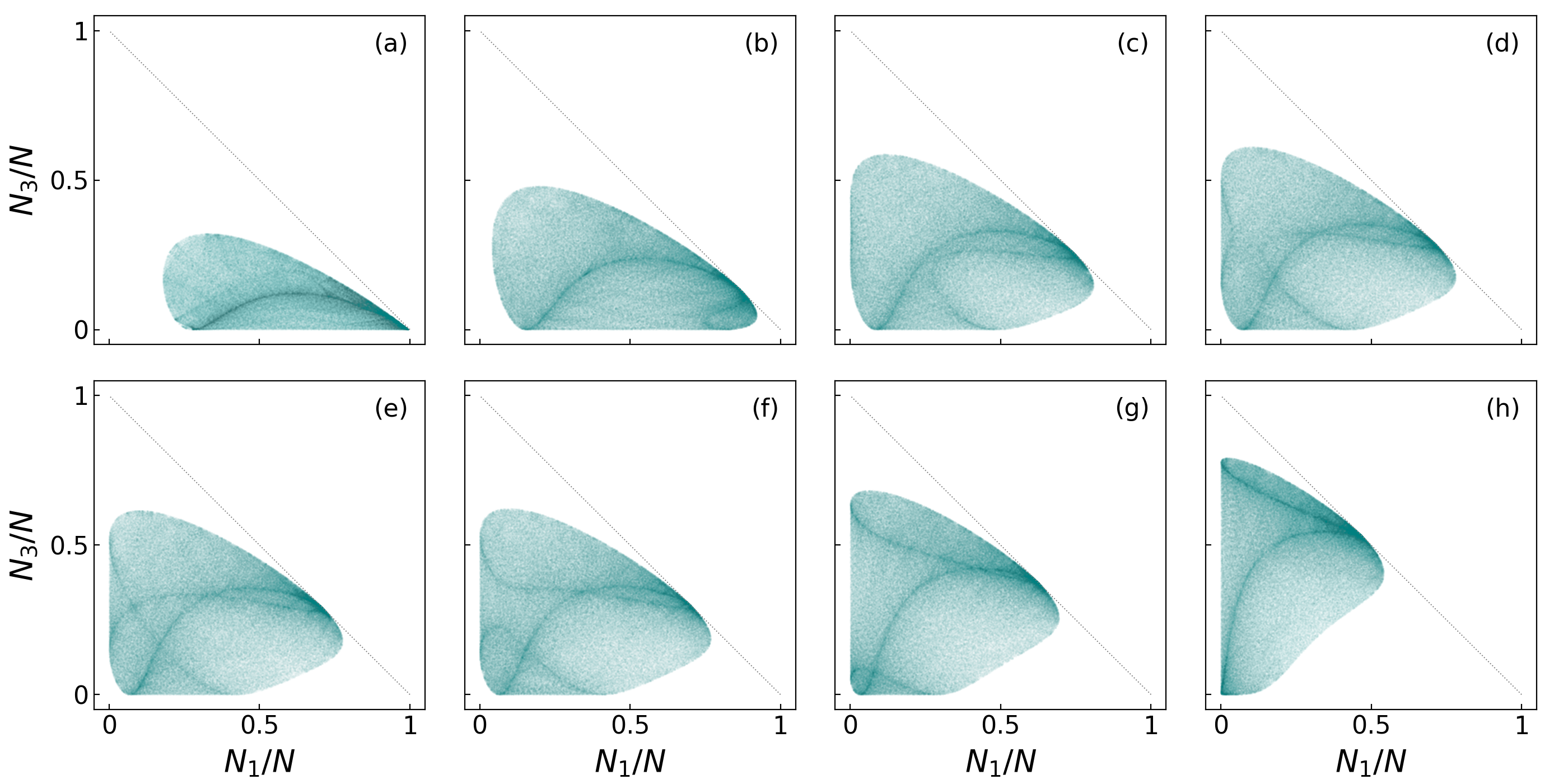}
    \caption{
    Representation of classical trajectories of \({\mathcal{H}}_{\text{cl}}\) in coordinates \(N_1\) and \(N_3\) for \(\epsilon_c = 1.5\). Each panel shows trajectories corresponding to initial conditions at specific energies, presented in the following sequence: \(E_{\text{classic}} \simeq -0.9\) (a),\; \(-0.4\) (b),\; \(-0.03\) (c),\; \(0.06\) (d), \(0.075\) (e),\; \(0.1\) (f),\;\(0.3\) (g)\; and \(0.8\) (h), the same as in  Fig. \ref{fig:QHeps15} and those marked in Fig. \ref{fig:PR}.  In panel (a), a superposition of multiple trajectories is presented, while panels (b-h) depict the evolution of a single initial condition for each energy. Panel (e) specifically illustrates the trajectory of the classical critical point \(P_c\), associated with \(E_c\). The correspondence between these classical trajectories and the Husimi projections in Fig. \ref{fig:QHeps15} is clearly visible.}
    \label{fig:Classicevol2}
\end{figure}

Fig. \ref{fig:QHeps15} presents the Husimi functions for sets of eigenvectors from different regions of the chaotic Hamiltonian spectrum. Each panel shows the projection of the average value of 200 eigenvectors, with eigenvalues centered around one of the energies marked in Fig. \ref{fig:PR}. In particular, Fig. \ref{fig:QHeps15}(e) represents the eigenvectors with eigenvalues centered at the critical energy \(E_{c}\).
In Fig. \ref{fig:Classicevol2}, we examine the classical model \({\mathcal{H}}_{\text{cl}}\) under the same chaotic parameters used in Fig. \ref{fig:QHeps15}.
The analysis is performed for various classical energies, \(E_{\text{classic}}\), corresponding to the values indicated in Fig. \ref{fig:PR}.
Each panel shows the trajectory corresponding to one of these energies. In panel (a), multiple initial conditions were evolved, while in the other cases, only a single trajectory is represented. This distinction arises because the energy in Fig. \ref{fig:Classicevol2}(a) lies in the region of the quantum spectrum where the PR is regular and localized (see Fig. 4), which leads to localized classical trajectories, unlike the other cases that lie in the chaotic region of the spectrum. Fig. \ref{fig:Classicevol2}(e) illustrates the trajectory of the critical point \(P_c\), corresponding to the critical energy \(E_c\).

%-------------------------------------------%

\onecolumngrid

%------------------------------------------%

\twocolumngrid

%-------------------------------------------%
%-------------------------------------------%
We can observe that the critical energy \(E_c\), associated with the intersecting trajectory loops (Figs. \ref{fig:QHeps15}(e) and \ref{fig:Classicevol2}(e)), is a transition energy between two distinct localization regimes. For this energy, any initial condition would yield the same trajectory.
There is a clear correspondence between the quantum and classical trajectories shown in Figs. \ref{fig:QHeps15} and \ref{fig:Classicevol2}, respectively. 
These trajectories also align with the participation ratio depicted in Fig. \ref{fig:PR}.

%-------------------------------------------%

\subsection{\label{sec:special}Special cases}
\begin{figure}[h!]
    \centering
        \includegraphics[width=0.9\linewidth]{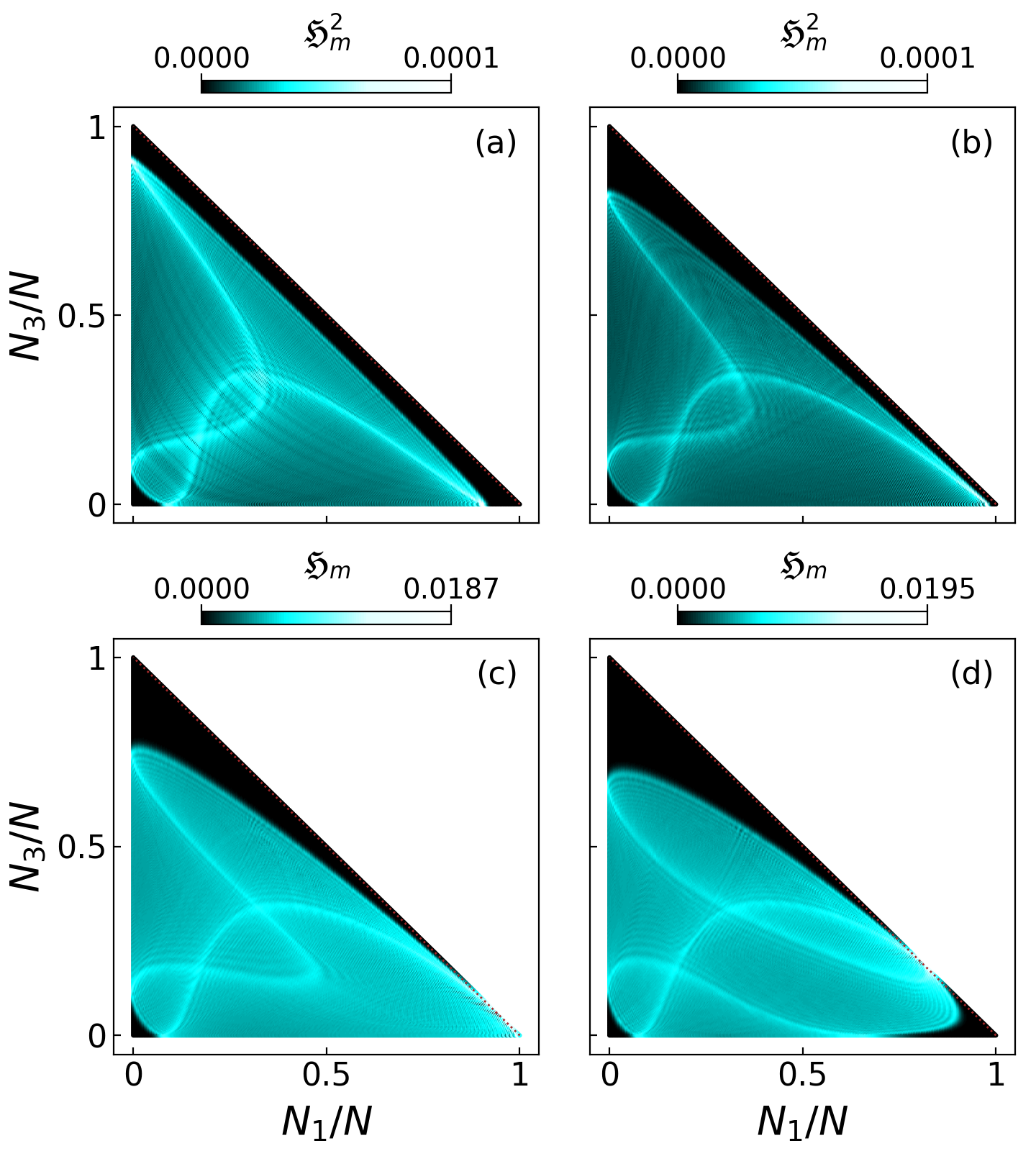}
    \caption{Representation of the quantum Hamiltonian \(\hat{H}\) as it varies with \(\epsilon\) using the Husimi function \(\mathfrak{H}_{E}(N_1, N_3)\) for \(N = 300\). Panels show the distribution of the average value of 200 eigenvectors at energy \(E_c \approx 0.075\) for \(\epsilon = 0\) (a), \(\epsilon = 0.4\) (b), \(\epsilon = 0.7\) (c), and \(\epsilon = 1\) (d). Unlike Fig. 2, where plots are centered at the critical energy for each \(\epsilon\), all plots here are centered at the same energy \(E_c\).}
    \label{fig:shrimp3}
\end{figure}

\begin{figure}[h]
    \centering
         \includegraphics[width=0.9\linewidth]{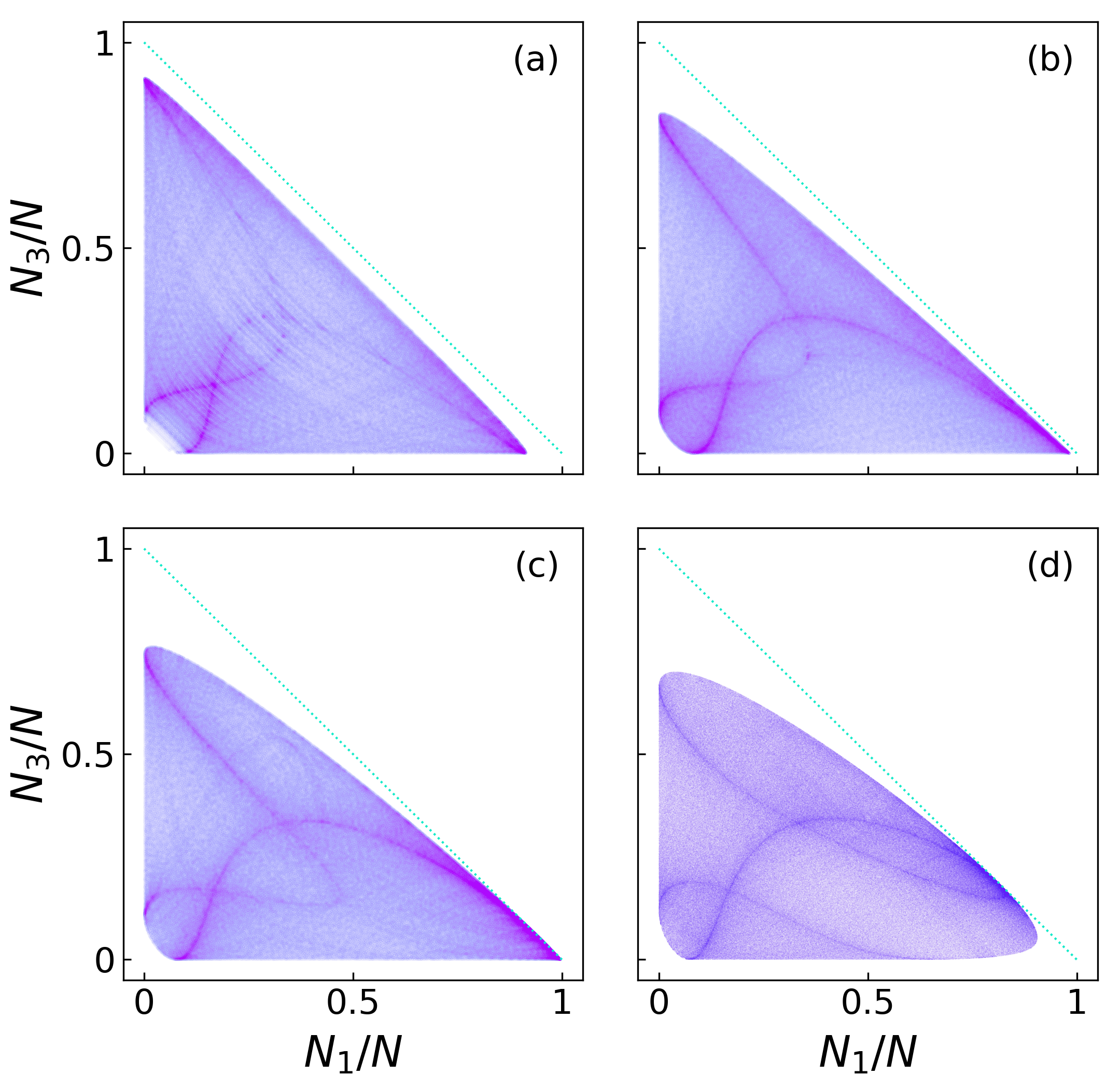}
          \caption{Classical Hamiltonian trajectories for various \(\epsilon\) values, all at energy \(E_c \approx 0.075\) are shown. Panels display trajectories for \(\epsilon = 0\) (a), \(\epsilon = 0.4\) (b), \(\epsilon = 0.7\) (c), and \(\epsilon = 1\) (d), as in Fig. \ref{fig:shrimp3}. The correspondence with Husimi projections in Fig. \ref{fig:shrimp3} is clear.}
    \label{fig:shrimp3cl}
\end{figure}

The patterns formed by the projections and  trajectories shown in the previous sections subtly reminded us of the shrimp shapes described by J.A.C. Gallas. Notably, there are some more unusual trajectories that reflect Gallas's patterns even more closely. These special trajectories are shown in Fig. \ref{fig:shrimp3} (Fig. \ref{fig:shrimp3cl}). They appear when analyzing $\hat{H}$ (\({\mathcal{H}}_{\text{cl}}\) ) for different values of \(\epsilon\), keeping the energy fixed at \(E_c\). Each panel in Fig. \ref{fig:shrimp3} displays the projection of 200 eigenstates with energies close to \(E_c\) for a given \(\epsilon<1.5\).

Interestingly, Fig. \ref{fig:shrimp3}(a), which corresponds to the integrable model (\(\epsilon = 0\)), closely resembles the Gallas patterns.
Additionally, we observe that the symmetry of the initial shape is distorted as \(\epsilon\) increases, while the overall structure of the curve appears to be preserved. This suggests that the kinks in the trajectories reflect variations in the parameter \(\epsilon\), for fixed energies. 

The classical counterparts of the figures in Fig. \ref{fig:shrimp3} are shown in Fig. \ref{fig:shrimp3cl}. Each panel in Fig. \ref{fig:shrimp3cl} presents a superposition of several trajectories associated with the value of $\epsilon$, but all at the same energy $E_c$.

%%%%%%%%%%%%%%%%%%%%%%%%%%%%%%%%%%%%%%%%%%%%
%%%%%%%%%%%%%%%%%%%%%%%%%%%%%%%%%%%%%%%%%%%%

\section{\label{discussion}Discussion}

We have analyzed the quantum-classical behavior of
a Hamiltonian that models ultracold dipolar atoms distributed in three aligned potential wells, when its integrability is broken. This comparison was made visually by examining the Husimi functions projected onto the \(N_1\) and \(N_3\) operators and comparing these with classical trajectories projected onto the \(N_1\) and \(N_3\) coordinates.

In Sec. \ref{sec:inttochaos}, we examined how the trajectories transform during the transition from the integrable system (\(\epsilon = 0\)) to a chaotic system (\(\epsilon_c = 1.5\)), and then to a localized system (\(\epsilon = 30\)). The transition between these regimes became evident. Initially, a diagonally symmetric trajectory stands out, close to \(N_2 = N\), reflecting the symmetry between wells 1 and 3 in the integrable Hamiltonian. This trajectory deforms as it approaches the critical parameter \(\epsilon_c = 1.5\), then begins to localize until it is reduced to a transverse symmetrical trajectory characterized by \(N_1 = N_3\), which tends towards isolated islands. 
This behavior is consistently observed in both quantum and classical systems. 
For the regular dynamics multiple classical trajectories were required to obtain the corresponding figures, and even then, the condensations were not perfectly equivalent. For the chaotic regime, a single trajectory was sufficient to generate a strongly matching pattern. Interestingly, the integrable and self-trapped models also exhibit clear classical-quantum correspondence.

Similar patterns emerged when we analyzed the Hamiltonian in the chaotic regime (Sec. \ref{sec:chaos}). Using the same parameters \(U\), \(J\), and \(\epsilon_c\) for both quantum and classical models, we compared quantum projections generated by subsets of eigenstates with classical trajectories, with both the quantum eigenvalues and classical initial conditions centered around the same energy ($E_m \simeq E_{classic}$). We observed that the system tends to localize at \(N_1\) for lower (negative) energies and at \(N_3\) for higher energies, with \(E_c\) marking the transition energy between these two localization regimes, where the trajectories intersect.

The visual correspondence between the quantum and classical systems in the chaotic regime is remarkably clear across the entire spectrum. As expected, it is much more precise and easier to obtain than in the analysis of the transition between the integrable, chaotic, and self-trapping regimes.
It may diminish somewhat for very low (or very high) eigenvalues located in the regular tails of the PR distribution, where it becomes more challenging to find classical trajectories that replicate the quantum average patterns over long periods, often requiring the superposition of multiple trajectories.

Quantum-classical correspondences generally become more evident in the spectrum of maximally chaotic parameters. Although chaotic regions do not strictly represent equilibrium systems, they can exhibit equilibrium-like statistical properties, such as ergodicity~\cite{BGC1984}. This behavior is verified here. The most faithful correspondences occur within the chaotic regime, particularly for parameters close to the unstable critical point and its associated critical energy. Additionally, strong correspondences were also observed outside the chaotic regime. The chaotic system's behavior, as inferred from the PR graph, is reflected in both quantum and classical systems.

The striking similarity between the phase-space mean projections of classical trajectories and those of Husimi distributions evokes the Principle of Uniform Semiclassical Condensation (PUSC) of Wigner eigenstate functions ~\cite{Robnik2020}. 

Remarkably, the evolution of these trajectory shapes unexpectedly unveiled "shrimp" patterns, especially those discussed in Sec. \ref{sec:special}, and encouraged us to present this work as a tribute to J.A.C. Gallas.

\section{acknowledgment}
K.W.W. and A.F. acknowledge financial support from the State of Rio Grande do Sul through FAPERGS - Edital FAPERGS/CNPq 07/2022 - Programa de Apoio à Fixação de Jovens Doutores no Brasil, contract 23/2551-0001836-5. A.F. acknowledges
support from CNPq (Conselho Nacional de Desenvolvimento Científico e Tecnológico) - Edital Universal 406563/2021-7. 
E.C. thanks the Brazilian CNPq agency for partial financial support. I.R. also thanks CNPq for partial support through contract 311876/2021-8. J.G.H acknowledges partial financial support from project PAPIIT-UNAM IN109523. The authors have no conflicts to disclose.

%%%%%%%%%%%%%%%%%%%%%%%%%%%%%%%%%%%%%%%%%%%%
%%%%%%%%%%%%%%%%%%%%%%%%%%%%%%%%%%%%%%%%%%%%

\appendix

\section{Participation Ratio (PR)}\label{app:PR}
\begin{figure}[h]
    \centering
  \includegraphics[width=1\linewidth]{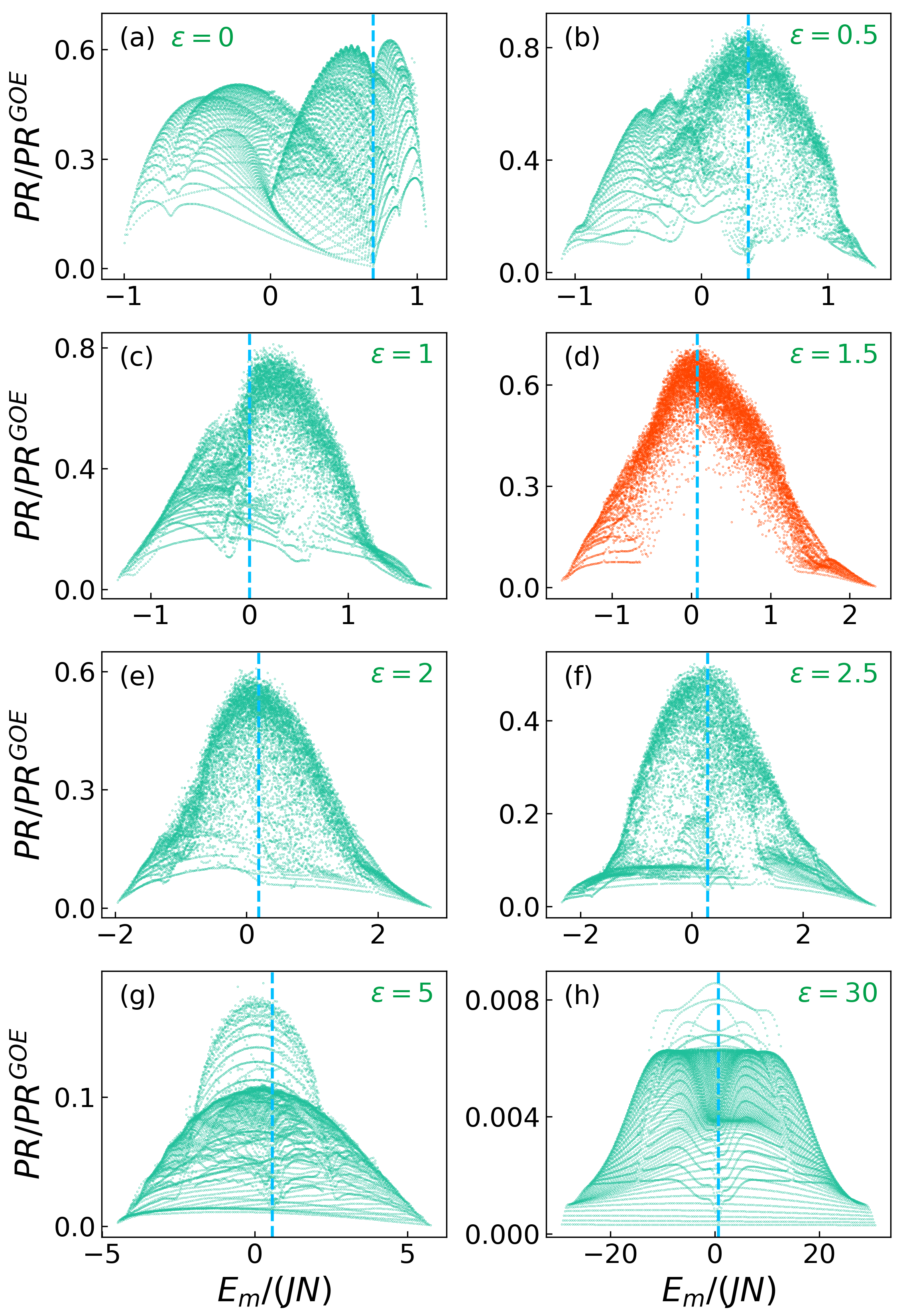}
    \caption{Participation ratio as a function of energy for the Hamiltonian \eqref{QH} for \(N = 150\), \(U = 0.7\), and \(J = 1\), with \(\epsilon\) values ranging from (a) \(\epsilon = 0\), (b) \(\epsilon = 0.5\), (c) \(\epsilon = 1\), (d) \(\epsilon = 1.5\), (e) \(\epsilon = 2\), (f) \(\epsilon = 2.5\), (g) \(\epsilon = 5\), and (h) \(\epsilon = 30\). . The dashed vertical lines represent the unstable critical energies derived from the classical model, which were considered in Figs. \ref{fig:Hevol} and \ref{fig:Classicevol}. 
    The chaotic case studied in Sec. \ref{sec:chaos} is highlighted in orange, panel (d). A meaningful comparison can be made between the cases in this figure and those presented in Figs. \ref{fig:Hevol} and \ref{fig:Classicevol}.}
    \label{fig:multiPR}
\end{figure}

The participation ratios of $\hat{H}$ for the various values of \(\epsilon\) discussed in Figs. \ref{fig:Hevol} and \ref{fig:Classicevol} are shown in Fig. \ref{fig:multiPR}. The classical critical energies for each case are indicated by dashed vertical lines. 

In PR distributions shown in Fig. \ref{fig:multiPR}, we observe the regular distribution of the PR of the eigenstates as a function of their energy, characteristic of integrable models, where the points can be visually connected along smooth lines, as seen in Fig. \ref{fig:Hevol}(a). At the same time, for each energy region the  dispersion of the PR values is large.
Approaching the chaotic domain, represented in Figs. \ref{fig:Hevol}(b-f), the patterns becomes less and less identifiable and the PR values have smaller dispersion, reaching a minimum at \(\epsilon = \epsilon_c\).  
The lost of patterns occurs primarily around the critical unstable energy of each spectrum, which is the focus of the analyses in Figs. \ref{fig:Hevol} and \ref{fig:Classicevol}. For \(\epsilon \gg \epsilon_c\), new lines with low PR emerge, clearly indicating localization, consistent with the results shown in Figs. \ref{fig:Hevol}(g-h) and \ref{fig:Classicevol}(g-h).
 Also, the PR values become more dispersed, but tend to localize as the participation ratio decreases. For \(\epsilon \gg \epsilon_c\), the system approaches a symmetric distribution around the critical energy, with low PR.
A more detailed analysis of the model in the chaotic limit can be found in Refs.~\cite{Wittmann2022, CastroPRA2024}.

%%%%%%%%%%%%%%%%%%%%%%%%%%%%%%%%%%%%%%%%%
%%%%%%%%%%%%%%%%%%%%%%%%%%%%%%%%%%%%%%%%%

%\nocite{*}
%\bibliography{biblio}

\begin{thebibliography}{41}%
\makeatletter
\providecommand \@ifxundefined [1]{%
 \@ifx{#1\undefined}
}%
\providecommand \@ifnum [1]{%
 \ifnum #1\expandafter \@firstoftwo
 \else \expandafter \@secondoftwo
 \fi
}%
\providecommand \@ifx [1]{%
 \ifx #1\expandafter \@firstoftwo
 \else \expandafter \@secondoftwo
 \fi
}%
\providecommand \natexlab [1]{#1}%
\providecommand \enquote  [1]{``#1''}%
\providecommand \bibnamefont  [1]{#1}%
\providecommand \bibfnamefont [1]{#1}%
\providecommand \citenamefont [1]{#1}%
\providecommand \href@noop [0]{\@secondoftwo}%
\providecommand \href [0]{\begingroup \@sanitize@url \@href}%
\providecommand \@href[1]{\@@startlink{#1}\@@href}%
\providecommand \@@href[1]{\endgroup#1\@@endlink}%
\providecommand \@sanitize@url [0]{\catcode `\\12\catcode `\$12\catcode `\&12\catcode `\#12\catcode `\^12\catcode `\_12\catcode `\%12\relax}%
\providecommand \@@startlink[1]{}%
\providecommand \@@endlink[0]{}%
\providecommand \url  [0]{\begingroup\@sanitize@url \@url }%
\providecommand \@url [1]{\endgroup\@href {#1}{\urlprefix }}%
\providecommand \urlprefix  [0]{URL }%
\providecommand \Eprint [0]{\href }%
\providecommand \doibase [0]{http://dx.doi.org/}%
\providecommand \selectlanguage [0]{\@gobble}%
\providecommand \bibinfo  [0]{\@secondoftwo}%
\providecommand \bibfield  [0]{\@secondoftwo}%
\providecommand \translation [1]{[#1]}%
\providecommand \BibitemOpen [0]{}%
\providecommand \bibitemStop [0]{}%
\providecommand \bibitemNoStop [0]{.\EOS\space}%
\providecommand \EOS [0]{\spacefactor3000\relax}%
\providecommand \BibitemShut  [1]{\csname bibitem#1\endcsname}%
\let\auto@bib@innerbib\@empty
%</preamble>
\bibitem [{\citenamefont {Gallas}(1993)}]{GALLAS1993}%
  \BibitemOpen
  \bibfield  {author} {\bibinfo {author} {\bibfnamefont {J.~A.~C.}\ \bibnamefont {Gallas}},\ }\bibfield  {title} {\enquote {\bibinfo {title} {Structure of the parameter space of the h\'enon map},}\ }\href {\doibase 10.1103/PhysRevLett.70.2714} {\bibfield  {journal} {\bibinfo  {journal} {Phys. Rev. Lett.}\ }\textbf {\bibinfo {volume} {70}},\ \bibinfo {pages} {2714--2717} (\bibinfo {year} {1993})}\BibitemShut {NoStop}%
\bibitem [{\citenamefont {Gallas}(1994)}]{GALLAS1994}%
  \BibitemOpen
  \bibfield  {author} {\bibinfo {author} {\bibfnamefont {J.~A.}\ \bibnamefont {Gallas}},\ }\bibfield  {title} {\enquote {\bibinfo {title} {Dissecting shrimps: results for some one-dimensional physical models},}\ }\href {\doibase https://doi.org/10.1016/0378-4371(94)90174-0} {\bibfield  {journal} {\bibinfo  {journal} {Physica A: Statistical Mechanics and its Applications}\ }\textbf {\bibinfo {volume} {202}},\ \bibinfo {pages} {196--223} (\bibinfo {year} {1994})}\BibitemShut {NoStop}%
\bibitem [{\citenamefont {Hunt}\ \emph {et~al.}(1999)\citenamefont {Hunt}, \citenamefont {Gallas}, \citenamefont {Grebogi}, \citenamefont {Yorke},\ and\ \citenamefont {Koçak}}]{HuntGallas1999}%
  \BibitemOpen
  \bibfield  {author} {\bibinfo {author} {\bibfnamefont {B.~R.}\ \bibnamefont {Hunt}}, \bibinfo {author} {\bibfnamefont {J.~A.}\ \bibnamefont {Gallas}}, \bibinfo {author} {\bibfnamefont {C.}~\bibnamefont {Grebogi}}, \bibinfo {author} {\bibfnamefont {J.~A.}\ \bibnamefont {Yorke}}, \ and\ \bibinfo {author} {\bibfnamefont {H.}~\bibnamefont {Koçak}},\ }\bibfield  {title} {\enquote {\bibinfo {title} {Bifurcation rigidity},}\ }\href {\doibase https://doi.org/10.1016/S0167-2789(98)00201-2} {\bibfield  {journal} {\bibinfo  {journal} {Physica D: Nonlinear Phenomena}\ }\textbf {\bibinfo {volume} {129}},\ \bibinfo {pages} {35--56} (\bibinfo {year} {1999})}\BibitemShut {NoStop}%
\bibitem [{\citenamefont {Bonatto}, \citenamefont {Garreau},\ and\ \citenamefont {Gallas}(2005)}]{BonattoGallas2005}%
  \BibitemOpen
  \bibfield  {author} {\bibinfo {author} {\bibfnamefont {C.}~\bibnamefont {Bonatto}}, \bibinfo {author} {\bibfnamefont {J.~C.}\ \bibnamefont {Garreau}}, \ and\ \bibinfo {author} {\bibfnamefont {J.~A.~C.}\ \bibnamefont {Gallas}},\ }\bibfield  {title} {\enquote {\bibinfo {title} {Self-similarities in the frequency-amplitude space of a loss-modulated $\mathrm{CO}_{2}$ laser},}\ }\href {\doibase 10.1103/PhysRevLett.95.143905} {\bibfield  {journal} {\bibinfo  {journal} {Phys. Rev. Lett.}\ }\textbf {\bibinfo {volume} {95}},\ \bibinfo {pages} {143905} (\bibinfo {year} {2005})}\BibitemShut {NoStop}%
\bibitem [{\citenamefont {Bonatto}\ and\ \citenamefont {Gallas}(2008)}]{BonattoGallas2008}%
  \BibitemOpen
  \bibfield  {author} {\bibinfo {author} {\bibfnamefont {C.}~\bibnamefont {Bonatto}}\ and\ \bibinfo {author} {\bibfnamefont {J.~A.~C.}\ \bibnamefont {Gallas}},\ }\bibfield  {title} {\enquote {\bibinfo {title} {Periodicity hub and nested spirals in the phase diagram of a simple resistive circuit},}\ }\href {\doibase 10.1103/PhysRevLett.101.054101} {\bibfield  {journal} {\bibinfo  {journal} {Phys. Rev. Lett.}\ }\textbf {\bibinfo {volume} {101}},\ \bibinfo {pages} {054101} (\bibinfo {year} {2008})}\BibitemShut {NoStop}%
\bibitem [{\citenamefont {Lahaye}, \citenamefont {Pfau},\ and\ \citenamefont {Santos}(2010)}]{Lahaye2010}%
  \BibitemOpen
  \bibfield  {author} {\bibinfo {author} {\bibfnamefont {T.}~\bibnamefont {Lahaye}}, \bibinfo {author} {\bibfnamefont {T.}~\bibnamefont {Pfau}}, \ and\ \bibinfo {author} {\bibfnamefont {L.}~\bibnamefont {Santos}},\ }\bibfield  {title} {\enquote {\bibinfo {title} {{M}esoscopic {E}nsembles of {P}olar {B}osons in {T}riple-{W}ell {P}otentials},}\ }\href {\doibase 10.1103/PhysRevLett.104.170404} {\bibfield  {journal} {\bibinfo  {journal} {Phys. Rev. Lett.}\ }\textbf {\bibinfo {volume} {104}},\ \bibinfo {pages} {170404} (\bibinfo {year} {2010})}\BibitemShut {NoStop}%
\bibitem [{\citenamefont {Wilsmann}\ \emph {et~al.}(2018)\citenamefont {Wilsmann}, \citenamefont {Ymai}, \citenamefont {Tonel}, \citenamefont {Links},\ and\ \citenamefont {Foerster}}]{Wittmann2018}%
  \BibitemOpen
  \bibfield  {author} {\bibinfo {author} {\bibfnamefont {K.~W.}\ \bibnamefont {Wilsmann}}, \bibinfo {author} {\bibfnamefont {L.~H.}\ \bibnamefont {Ymai}}, \bibinfo {author} {\bibfnamefont {A.~P.}\ \bibnamefont {Tonel}}, \bibinfo {author} {\bibfnamefont {J.}~\bibnamefont {Links}}, \ and\ \bibinfo {author} {\bibfnamefont {A.}~\bibnamefont {Foerster}},\ }\bibfield  {title} {\enquote {\bibinfo {title} {{C}ontrol of tunneling in an atomtronic switching device},}\ }\href {\doibase 10.1038/s42005-018-0089-1} {\bibfield  {journal} {\bibinfo  {journal} {Comm. Phys.}\ }\textbf {\bibinfo {volume} {1}} (\bibinfo {year} {2018}),\ 10.1038/s42005-018-0089-1}\BibitemShut {NoStop}%
\bibitem [{\citenamefont {Ymai}\ \emph {et~al.}(2017)\citenamefont {Ymai}, \citenamefont {Tonel}, \citenamefont {Foerster},\ and\ \citenamefont {Links}}]{Ymai2017}%
  \BibitemOpen
  \bibfield  {author} {\bibinfo {author} {\bibfnamefont {L.~H.}\ \bibnamefont {Ymai}}, \bibinfo {author} {\bibfnamefont {A.~P.}\ \bibnamefont {Tonel}}, \bibinfo {author} {\bibfnamefont {A.}~\bibnamefont {Foerster}}, \ and\ \bibinfo {author} {\bibfnamefont {J.}~\bibnamefont {Links}},\ }\bibfield  {title} {\enquote {\bibinfo {title} {{Q}uantum integrable multi-well tunneling models},}\ }\href {\doibase 10.1088/1751-8121/aa7227} {\bibfield  {journal} {\bibinfo  {journal} {J. Phys. A}\ }\textbf {\bibinfo {volume} {50}},\ \bibinfo {pages} {264001} (\bibinfo {year} {2017})}\BibitemShut {NoStop}%
\bibitem [{\citenamefont {Tonel}, \citenamefont {Links},\ and\ \citenamefont {Foerster}(2005)}]{Tonel2005}%
  \BibitemOpen
  \bibfield  {author} {\bibinfo {author} {\bibfnamefont {A.~P.}\ \bibnamefont {Tonel}}, \bibinfo {author} {\bibfnamefont {J.}~\bibnamefont {Links}}, \ and\ \bibinfo {author} {\bibfnamefont {A.}~\bibnamefont {Foerster}},\ }\bibfield  {title} {\enquote {\bibinfo {title} {Quantum dynamics of a model for two {J}osephson-coupled {B}ose{\textendash}{E}instein condensates},}\ }\href {\doibase 10.1088/0305-4470/38/6/004} {\bibfield  {journal} {\bibinfo  {journal} {J. Phys. A}\ }\textbf {\bibinfo {volume} {38}},\ \bibinfo {pages} {1235--1245} (\bibinfo {year} {2005})}\BibitemShut {NoStop}%
\bibitem [{\citenamefont {Links}\ \emph {et~al.}(2006)\citenamefont {Links}, \citenamefont {Foerster}, \citenamefont {Tonel},\ and\ \citenamefont {Santos}}]{Links2006}%
  \BibitemOpen
  \bibfield  {author} {\bibinfo {author} {\bibfnamefont {J.}~\bibnamefont {Links}}, \bibinfo {author} {\bibfnamefont {A.}~\bibnamefont {Foerster}}, \bibinfo {author} {\bibfnamefont {A.}~\bibnamefont {Tonel}}, \ and\ \bibinfo {author} {\bibfnamefont {G.}~\bibnamefont {Santos}},\ }\bibfield  {title} {\enquote {\bibinfo {title} {The two-site {B}ose-{H}ubbard model},}\ }\href {\doibase https://doi.org/10.1007/s00023-006-0295-3} {\bibfield  {journal} {\bibinfo  {journal} {Ann. Henri Poincaré}\ }\textbf {\bibinfo {volume} {7}},\ \bibinfo {pages} {1591} (\bibinfo {year} {2006})}\BibitemShut {NoStop}%
\bibitem [{\citenamefont {Tonel}\ \emph {et~al.}(2020)\citenamefont {Tonel}, \citenamefont {Ymai}, \citenamefont {W.}, \citenamefont {Foerster},\ and\ \citenamefont {Links}}]{Tonel2020}%
  \BibitemOpen
  \bibfield  {author} {\bibinfo {author} {\bibfnamefont {A.~P.}\ \bibnamefont {Tonel}}, \bibinfo {author} {\bibfnamefont {L.~H.}\ \bibnamefont {Ymai}}, \bibinfo {author} {\bibfnamefont {K.~W.}\ \bibnamefont {W.}}, \bibinfo {author} {\bibfnamefont {A.}~\bibnamefont {Foerster}}, \ and\ \bibinfo {author} {\bibfnamefont {J.}~\bibnamefont {Links}},\ }\bibfield  {title} {\enquote {\bibinfo {title} {{Entangled states of dipolar bosons generated in a triple-well potential}},}\ }\href {\doibase 10.21468/SciPostPhysCore.2.1.003} {\bibfield  {journal} {\bibinfo  {journal} {SciPost Phys. Core}\ }\textbf {\bibinfo {volume} {2}},\ \bibinfo {pages} {3} (\bibinfo {year} {2020})}\BibitemShut {NoStop}%
\bibitem [{\citenamefont {W.}\ \emph {et~al.}(2023)\citenamefont {W.}, \citenamefont {Ymai}, \citenamefont {Barros}, \citenamefont {Links},\ and\ \citenamefont {Foerster}}]{wittmann2023}%
  \BibitemOpen
  \bibfield  {author} {\bibinfo {author} {\bibfnamefont {K.~W.}\ \bibnamefont {W.}}, \bibinfo {author} {\bibfnamefont {L.~H.}\ \bibnamefont {Ymai}}, \bibinfo {author} {\bibfnamefont {B.~H.~C.}\ \bibnamefont {Barros}}, \bibinfo {author} {\bibfnamefont {J.}~\bibnamefont {Links}}, \ and\ \bibinfo {author} {\bibfnamefont {A.}~\bibnamefont {Foerster}},\ }\bibfield  {title} {\enquote {\bibinfo {title} {Controlling entanglement in a triple-well system of dipolar atoms},}\ }\href {\doibase 10.1103/PhysRevA.108.033313} {\bibfield  {journal} {\bibinfo  {journal} {Phys. Rev. A}\ }\textbf {\bibinfo {volume} {108}},\ \bibinfo {pages} {033313} (\bibinfo {year} {2023})}\BibitemShut {NoStop}%
\bibitem [{\citenamefont {Gr\"un}\ \emph {et~al.}(2022)\citenamefont {Gr\"un}, \citenamefont {Ymai}, \citenamefont {Wittmann~W.}, \citenamefont {Tonel}, \citenamefont {Foerster},\ and\ \citenamefont {Links}}]{grun2022interfer}%
  \BibitemOpen
  \bibfield  {author} {\bibinfo {author} {\bibfnamefont {D.~S.}\ \bibnamefont {Gr\"un}}, \bibinfo {author} {\bibfnamefont {L.~H.}\ \bibnamefont {Ymai}}, \bibinfo {author} {\bibfnamefont {K.}~\bibnamefont {Wittmann~W.}}, \bibinfo {author} {\bibfnamefont {A.~P.}\ \bibnamefont {Tonel}}, \bibinfo {author} {\bibfnamefont {A.}~\bibnamefont {Foerster}}, \ and\ \bibinfo {author} {\bibfnamefont {J.}~\bibnamefont {Links}},\ }\bibfield  {title} {\enquote {\bibinfo {title} {Integrable atomtronic interferometry},}\ }\href {\doibase 10.1103/PhysRevLett.129.020401} {\bibfield  {journal} {\bibinfo  {journal} {Phys. Rev. Lett.}\ }\textbf {\bibinfo {volume} {129}},\ \bibinfo {pages} {020401} (\bibinfo {year} {2022})}\BibitemShut {NoStop}%
\bibitem [{\citenamefont {Gr{\"u}n}\ \emph {et~al.}(2022)\citenamefont {Gr{\"u}n}, \citenamefont {Wittmann~W.}, \citenamefont {Ymai}, \citenamefont {Links},\ and\ \citenamefont {Foerster}}]{grun2022protocol}%
  \BibitemOpen
  \bibfield  {author} {\bibinfo {author} {\bibfnamefont {D.~S.}\ \bibnamefont {Gr{\"u}n}}, \bibinfo {author} {\bibfnamefont {K.}~\bibnamefont {Wittmann~W.}}, \bibinfo {author} {\bibfnamefont {L.~H.}\ \bibnamefont {Ymai}}, \bibinfo {author} {\bibfnamefont {J.}~\bibnamefont {Links}}, \ and\ \bibinfo {author} {\bibfnamefont {A.}~\bibnamefont {Foerster}},\ }\bibfield  {title} {\enquote {\bibinfo {title} {Protocol designs for {NOON} states},}\ }\href {\doibase 10.1038/s42005-022-00812-7} {\bibfield  {journal} {\bibinfo  {journal} {Commun. Phys.}\ }\textbf {\bibinfo {volume} {5}},\ \bibinfo {pages} {36} (\bibinfo {year} {2022})}\BibitemShut {NoStop}%
\bibitem [{\citenamefont {Mi}\ \emph {et~al.}(2021)\citenamefont {Mi}, \citenamefont {Roushan}, \citenamefont {Quintana}, \citenamefont {Mandrà}, \citenamefont {Marshall}, \citenamefont {Neill}, \citenamefont {Arute}, \citenamefont {Arya}, \citenamefont {Atalaya}, \citenamefont {Babbush}, \citenamefont {Bardin}, \citenamefont {Barends}, \citenamefont {Basso}, \citenamefont {Bengtsson}, \citenamefont {Boixo}, \citenamefont {Bourassa}, \citenamefont {Broughton}, \citenamefont {Buckley}, \citenamefont {Buell}, \citenamefont {Burkett}, \citenamefont {Bushnell}, \citenamefont {Chen}, \citenamefont {Chiaro}, \citenamefont {Collins}, \citenamefont {Courtney}, \citenamefont {Demura}, \citenamefont {Derk}, \citenamefont {Dunsworth}, \citenamefont {Eppens}, \citenamefont {Erickson}, \citenamefont {Farhi}, \citenamefont {Fowler}, \citenamefont {Foxen}, \citenamefont {Gidney}, \citenamefont {Giustina}, \citenamefont {Gross}, \citenamefont {Harrigan}, \citenamefont {Harrington}, \citenamefont {Hilton}, \citenamefont
  {Ho}, \citenamefont {Hong}, \citenamefont {Huang}, \citenamefont {Huggins}, \citenamefont {Ioffe}, \citenamefont {Isakov}, \citenamefont {Jeffrey}, \citenamefont {Jiang}, \citenamefont {Jones}, \citenamefont {Kafri}, \citenamefont {Kelly}, \citenamefont {Kim}, \citenamefont {Kitaev}, \citenamefont {Klimov}, \citenamefont {Korotkov}, \citenamefont {Kostritsa}, \citenamefont {Landhuis}, \citenamefont {Laptev}, \citenamefont {Lucero}, \citenamefont {Martin}, \citenamefont {McClean}, \citenamefont {McCourt}, \citenamefont {McEwen}, \citenamefont {Megrant}, \citenamefont {Miao}, \citenamefont {Mohseni}, \citenamefont {Montazeri}, \citenamefont {Mruczkiewicz}, \citenamefont {Mutus}, \citenamefont {Naaman}, \citenamefont {Neeley}, \citenamefont {Newman}, \citenamefont {Niu}, \citenamefont {O’Brien}, \citenamefont {Opremcak}, \citenamefont {Ostby}, \citenamefont {Pato}, \citenamefont {Petukhov}, \citenamefont {Redd}, \citenamefont {Rubin}, \citenamefont {Sank}, \citenamefont {Satzinger}, \citenamefont {Shvarts},
  \citenamefont {Strain}, \citenamefont {Szalay}, \citenamefont {Trevithick}, \citenamefont {Villalonga}, \citenamefont {White}, \citenamefont {Yao}, \citenamefont {Yeh}, \citenamefont {Zalcman}, \citenamefont {Neven}, \citenamefont {Aleiner}, \citenamefont {Kechedzhi}, \citenamefont {Smelyanskiy},\ and\ \citenamefont {Chen}}]{Marissa2021}%
  \BibitemOpen
  \bibfield  {author} {\bibinfo {author} {\bibfnamefont {X.}~\bibnamefont {Mi}}, \bibinfo {author} {\bibfnamefont {P.}~\bibnamefont {Roushan}}, \bibinfo {author} {\bibfnamefont {C.}~\bibnamefont {Quintana}}, \bibinfo {author} {\bibfnamefont {S.}~\bibnamefont {Mandrà}}, \bibinfo {author} {\bibfnamefont {J.}~\bibnamefont {Marshall}}, \bibinfo {author} {\bibfnamefont {C.}~\bibnamefont {Neill}}, \bibinfo {author} {\bibfnamefont {F.}~\bibnamefont {Arute}}, \bibinfo {author} {\bibfnamefont {K.}~\bibnamefont {Arya}}, \bibinfo {author} {\bibfnamefont {J.}~\bibnamefont {Atalaya}}, \bibinfo {author} {\bibfnamefont {R.}~\bibnamefont {Babbush}}, \bibinfo {author} {\bibfnamefont {J.~C.}\ \bibnamefont {Bardin}}, \bibinfo {author} {\bibfnamefont {R.}~\bibnamefont {Barends}}, \bibinfo {author} {\bibfnamefont {J.}~\bibnamefont {Basso}}, \bibinfo {author} {\bibfnamefont {A.}~\bibnamefont {Bengtsson}}, \bibinfo {author} {\bibfnamefont {S.}~\bibnamefont {Boixo}}, \bibinfo {author} {\bibfnamefont {A.}~\bibnamefont {Bourassa}},
  \bibinfo {author} {\bibfnamefont {M.}~\bibnamefont {Broughton}}, \bibinfo {author} {\bibfnamefont {B.~B.}\ \bibnamefont {Buckley}}, \bibinfo {author} {\bibfnamefont {D.~A.}\ \bibnamefont {Buell}}, \bibinfo {author} {\bibfnamefont {B.}~\bibnamefont {Burkett}}, \bibinfo {author} {\bibfnamefont {N.}~\bibnamefont {Bushnell}}, \bibinfo {author} {\bibfnamefont {Z.}~\bibnamefont {Chen}}, \bibinfo {author} {\bibfnamefont {B.}~\bibnamefont {Chiaro}}, \bibinfo {author} {\bibfnamefont {R.}~\bibnamefont {Collins}}, \bibinfo {author} {\bibfnamefont {W.}~\bibnamefont {Courtney}}, \bibinfo {author} {\bibfnamefont {S.}~\bibnamefont {Demura}}, \bibinfo {author} {\bibfnamefont {A.~R.}\ \bibnamefont {Derk}}, \bibinfo {author} {\bibfnamefont {A.}~\bibnamefont {Dunsworth}}, \bibinfo {author} {\bibfnamefont {D.}~\bibnamefont {Eppens}}, \bibinfo {author} {\bibfnamefont {C.}~\bibnamefont {Erickson}}, \bibinfo {author} {\bibfnamefont {E.}~\bibnamefont {Farhi}}, \bibinfo {author} {\bibfnamefont {A.~G.}\ \bibnamefont {Fowler}},
  \bibinfo {author} {\bibfnamefont {B.}~\bibnamefont {Foxen}}, \bibinfo {author} {\bibfnamefont {C.}~\bibnamefont {Gidney}}, \bibinfo {author} {\bibfnamefont {M.}~\bibnamefont {Giustina}}, \bibinfo {author} {\bibfnamefont {J.~A.}\ \bibnamefont {Gross}}, \bibinfo {author} {\bibfnamefont {M.~P.}\ \bibnamefont {Harrigan}}, \bibinfo {author} {\bibfnamefont {S.~D.}\ \bibnamefont {Harrington}}, \bibinfo {author} {\bibfnamefont {J.}~\bibnamefont {Hilton}}, \bibinfo {author} {\bibfnamefont {A.}~\bibnamefont {Ho}}, \bibinfo {author} {\bibfnamefont {S.}~\bibnamefont {Hong}}, \bibinfo {author} {\bibfnamefont {T.}~\bibnamefont {Huang}}, \bibinfo {author} {\bibfnamefont {W.~J.}\ \bibnamefont {Huggins}}, \bibinfo {author} {\bibfnamefont {L.~B.}\ \bibnamefont {Ioffe}}, \bibinfo {author} {\bibfnamefont {S.~V.}\ \bibnamefont {Isakov}}, \bibinfo {author} {\bibfnamefont {E.}~\bibnamefont {Jeffrey}}, \bibinfo {author} {\bibfnamefont {Z.}~\bibnamefont {Jiang}}, \bibinfo {author} {\bibfnamefont {C.}~\bibnamefont {Jones}}, \bibinfo
  {author} {\bibfnamefont {D.}~\bibnamefont {Kafri}}, \bibinfo {author} {\bibfnamefont {J.}~\bibnamefont {Kelly}}, \bibinfo {author} {\bibfnamefont {S.}~\bibnamefont {Kim}}, \bibinfo {author} {\bibfnamefont {A.}~\bibnamefont {Kitaev}}, \bibinfo {author} {\bibfnamefont {P.~V.}\ \bibnamefont {Klimov}}, \bibinfo {author} {\bibfnamefont {A.~N.}\ \bibnamefont {Korotkov}}, \bibinfo {author} {\bibfnamefont {F.}~\bibnamefont {Kostritsa}}, \bibinfo {author} {\bibfnamefont {D.}~\bibnamefont {Landhuis}}, \bibinfo {author} {\bibfnamefont {P.}~\bibnamefont {Laptev}}, \bibinfo {author} {\bibfnamefont {E.}~\bibnamefont {Lucero}}, \bibinfo {author} {\bibfnamefont {O.}~\bibnamefont {Martin}}, \bibinfo {author} {\bibfnamefont {J.~R.}\ \bibnamefont {McClean}}, \bibinfo {author} {\bibfnamefont {T.}~\bibnamefont {McCourt}}, \bibinfo {author} {\bibfnamefont {M.}~\bibnamefont {McEwen}}, \bibinfo {author} {\bibfnamefont {A.}~\bibnamefont {Megrant}}, \bibinfo {author} {\bibfnamefont {K.~C.}\ \bibnamefont {Miao}}, \bibinfo {author}
  {\bibfnamefont {M.}~\bibnamefont {Mohseni}}, \bibinfo {author} {\bibfnamefont {S.}~\bibnamefont {Montazeri}}, \bibinfo {author} {\bibfnamefont {W.}~\bibnamefont {Mruczkiewicz}}, \bibinfo {author} {\bibfnamefont {J.}~\bibnamefont {Mutus}}, \bibinfo {author} {\bibfnamefont {O.}~\bibnamefont {Naaman}}, \bibinfo {author} {\bibfnamefont {M.}~\bibnamefont {Neeley}}, \bibinfo {author} {\bibfnamefont {M.}~\bibnamefont {Newman}}, \bibinfo {author} {\bibfnamefont {M.~Y.}\ \bibnamefont {Niu}}, \bibinfo {author} {\bibfnamefont {T.~E.}\ \bibnamefont {O’Brien}}, \bibinfo {author} {\bibfnamefont {A.}~\bibnamefont {Opremcak}}, \bibinfo {author} {\bibfnamefont {E.}~\bibnamefont {Ostby}}, \bibinfo {author} {\bibfnamefont {B.}~\bibnamefont {Pato}}, \bibinfo {author} {\bibfnamefont {A.}~\bibnamefont {Petukhov}}, \bibinfo {author} {\bibfnamefont {N.}~\bibnamefont {Redd}}, \bibinfo {author} {\bibfnamefont {N.~C.}\ \bibnamefont {Rubin}}, \bibinfo {author} {\bibfnamefont {D.}~\bibnamefont {Sank}}, \bibinfo {author}
  {\bibfnamefont {K.~J.}\ \bibnamefont {Satzinger}}, \bibinfo {author} {\bibfnamefont {V.}~\bibnamefont {Shvarts}}, \bibinfo {author} {\bibfnamefont {D.}~\bibnamefont {Strain}}, \bibinfo {author} {\bibfnamefont {M.}~\bibnamefont {Szalay}}, \bibinfo {author} {\bibfnamefont {M.~D.}\ \bibnamefont {Trevithick}}, \bibinfo {author} {\bibfnamefont {B.}~\bibnamefont {Villalonga}}, \bibinfo {author} {\bibfnamefont {T.}~\bibnamefont {White}}, \bibinfo {author} {\bibfnamefont {Z.~J.}\ \bibnamefont {Yao}}, \bibinfo {author} {\bibfnamefont {P.}~\bibnamefont {Yeh}}, \bibinfo {author} {\bibfnamefont {A.}~\bibnamefont {Zalcman}}, \bibinfo {author} {\bibfnamefont {H.}~\bibnamefont {Neven}}, \bibinfo {author} {\bibfnamefont {I.}~\bibnamefont {Aleiner}}, \bibinfo {author} {\bibfnamefont {K.}~\bibnamefont {Kechedzhi}}, \bibinfo {author} {\bibfnamefont {V.}~\bibnamefont {Smelyanskiy}}, \ and\ \bibinfo {author} {\bibfnamefont {Y.}~\bibnamefont {Chen}},\ }\bibfield  {title} {\enquote {\bibinfo {title} {Information scrambling in
  quantum circuits},}\ }\href {\doibase 10.1126/science.abg5029} {\bibfield  {journal} {\bibinfo  {journal} {Science}\ }\textbf {\bibinfo {volume} {374}},\ \bibinfo {pages} {1479--1483} (\bibinfo {year} {2021})},\ \Eprint {http://arxiv.org/abs/https://www.science.org/doi/pdf/10.1126/science.abg5029} {https://www.science.org/doi/pdf/10.1126/science.abg5029} \BibitemShut {NoStop}%
\bibitem [{\citenamefont {Castro}\ \emph {et~al.}(2021)\citenamefont {Castro}, \citenamefont {Ch{\'{a}}vez-Carlos}, \citenamefont {Roditi}, \citenamefont {Santos},\ and\ \citenamefont {Hirsch}}]{Castro21}%
  \BibitemOpen
  \bibfield  {author} {\bibinfo {author} {\bibfnamefont {E.~R.}\ \bibnamefont {Castro}}, \bibinfo {author} {\bibfnamefont {J.}~\bibnamefont {Ch{\'{a}}vez-Carlos}}, \bibinfo {author} {\bibfnamefont {I.}~\bibnamefont {Roditi}}, \bibinfo {author} {\bibfnamefont {L.~F.}\ \bibnamefont {Santos}}, \ and\ \bibinfo {author} {\bibfnamefont {J.~G.}\ \bibnamefont {Hirsch}},\ }\bibfield  {title} {\enquote {\bibinfo {title} {Quantum-classical correspondence of a system of interacting bosons in a triple-well potential},}\ }\href {\doibase 10.22331/q-2021-10-19-563} {\bibfield  {journal} {\bibinfo  {journal} {{Quantum}}\ }\textbf {\bibinfo {volume} {5}},\ \bibinfo {pages} {563} (\bibinfo {year} {2021})}\BibitemShut {NoStop}%
\bibitem [{\citenamefont {Wittmann~W.}\ \emph {et~al.}(2022)\citenamefont {Wittmann~W.}, \citenamefont {Castro}, \citenamefont {Foerster},\ and\ \citenamefont {Santos}}]{Wittmann2022}%
  \BibitemOpen
  \bibfield  {author} {\bibinfo {author} {\bibfnamefont {K.}~\bibnamefont {Wittmann~W.}}, \bibinfo {author} {\bibfnamefont {E.~R.}\ \bibnamefont {Castro}}, \bibinfo {author} {\bibfnamefont {A.}~\bibnamefont {Foerster}}, \ and\ \bibinfo {author} {\bibfnamefont {L.~F.}\ \bibnamefont {Santos}},\ }\bibfield  {title} {\enquote {\bibinfo {title} {Interacting bosons in a triple well: Preface of many-body quantum chaos},}\ }\href {\doibase 10.1103/PhysRevE.105.034204} {\bibfield  {journal} {\bibinfo  {journal} {Phys. Rev. E}\ }\textbf {\bibinfo {volume} {105}},\ \bibinfo {pages} {034204} (\bibinfo {year} {2022})}\BibitemShut {NoStop}%
\bibitem [{\citenamefont {Castro}\ \emph {et~al.}(2024)\citenamefont {Castro}, \citenamefont {W.}, \citenamefont {Ch\'avez-Carlos}, \citenamefont {Roditi}, \citenamefont {Foerster},\ and\ \citenamefont {Hirsch}}]{CastroPRA2024}%
  \BibitemOpen
  \bibfield  {author} {\bibinfo {author} {\bibfnamefont {E.~R.}\ \bibnamefont {Castro}}, \bibinfo {author} {\bibfnamefont {K.~W.}\ \bibnamefont {W.}}, \bibinfo {author} {\bibfnamefont {J.}~\bibnamefont {Ch\'avez-Carlos}}, \bibinfo {author} {\bibfnamefont {I.}~\bibnamefont {Roditi}}, \bibinfo {author} {\bibfnamefont {A.}~\bibnamefont {Foerster}}, \ and\ \bibinfo {author} {\bibfnamefont {J.~G.}\ \bibnamefont {Hirsch}},\ }\bibfield  {title} {\enquote {\bibinfo {title} {Quantum-classical correspondence in a triple-well bosonic model: From integrability to chaos},}\ }\href {\doibase 10.1103/PhysRevA.109.032225} {\bibfield  {journal} {\bibinfo  {journal} {Phys. Rev. A}\ }\textbf {\bibinfo {volume} {109}},\ \bibinfo {pages} {032225} (\bibinfo {year} {2024})}\BibitemShut {NoStop}%
\bibitem [{\citenamefont {Nemoto}\ \emph {et~al.}(2000)\citenamefont {Nemoto}, \citenamefont {Holmes}, \citenamefont {Milburn},\ and\ \citenamefont {Munro}}]{Nemoto2000}%
  \BibitemOpen
  \bibfield  {author} {\bibinfo {author} {\bibfnamefont {K.}~\bibnamefont {Nemoto}}, \bibinfo {author} {\bibfnamefont {C.~A.}\ \bibnamefont {Holmes}}, \bibinfo {author} {\bibfnamefont {G.~J.}\ \bibnamefont {Milburn}}, \ and\ \bibinfo {author} {\bibfnamefont {W.~J.}\ \bibnamefont {Munro}},\ }\bibfield  {title} {\enquote {\bibinfo {title} {{Q}uantum dynamics of three coupled atomic {B}ose-{E}instein condensates},}\ }\href {\doibase 10.1103/PhysRevA.63.013604} {\bibfield  {journal} {\bibinfo  {journal} {Phys. Rev. A}\ }\textbf {\bibinfo {volume} {63}},\ \bibinfo {pages} {013604} (\bibinfo {year} {2000})}\BibitemShut {NoStop}%
\bibitem [{\citenamefont {Franzosi}\ and\ \citenamefont {Penna}(2001)}]{Franzosi2001}%
  \BibitemOpen
  \bibfield  {author} {\bibinfo {author} {\bibfnamefont {R.}~\bibnamefont {Franzosi}}\ and\ \bibinfo {author} {\bibfnamefont {V.}~\bibnamefont {Penna}},\ }\bibfield  {title} {\enquote {\bibinfo {title} {Self-trapping mechanisms in the dynamics of three coupled {B}ose-{E}instein condensates},}\ }\href {\doibase 10.1103/PhysRevA.65.013601} {\bibfield  {journal} {\bibinfo  {journal} {Phys. Rev. A}\ }\textbf {\bibinfo {volume} {65}},\ \bibinfo {pages} {013601} (\bibinfo {year} {2001})}\BibitemShut {NoStop}%
\bibitem [{\citenamefont {Mossmann}\ and\ \citenamefont {Jung}(2006)}]{Mossmann2006}%
  \BibitemOpen
  \bibfield  {author} {\bibinfo {author} {\bibfnamefont {S.}~\bibnamefont {Mossmann}}\ and\ \bibinfo {author} {\bibfnamefont {C.}~\bibnamefont {Jung}},\ }\bibfield  {title} {\enquote {\bibinfo {title} {Semiclassical approach to {B}ose-{E}instein condensates in a triple well potential},}\ }\href {\doibase 10.1103/PhysRevA.74.033601} {\bibfield  {journal} {\bibinfo  {journal} {Phys. Rev. A}\ }\textbf {\bibinfo {volume} {74}},\ \bibinfo {pages} {033601} (\bibinfo {year} {2006})}\BibitemShut {NoStop}%
\bibitem [{\citenamefont {Graefe}, \citenamefont {Korsch},\ and\ \citenamefont {Witthaut}(2006)}]{Graefe2006}%
  \BibitemOpen
  \bibfield  {author} {\bibinfo {author} {\bibfnamefont {E.~M.}\ \bibnamefont {Graefe}}, \bibinfo {author} {\bibfnamefont {H.~J.}\ \bibnamefont {Korsch}}, \ and\ \bibinfo {author} {\bibfnamefont {D.}~\bibnamefont {Witthaut}},\ }\bibfield  {title} {\enquote {\bibinfo {title} {Mean-field dynamics of a {B}ose-{E}instein condensate in a time-dependent triple-well trap: {N}onlinear eigenstates, {L}andau-{Z}ener models, and stimulated {R}aman adiabatic passage},}\ }\href {\doibase 10.1103/PhysRevA.73.013617} {\bibfield  {journal} {\bibinfo  {journal} {Phys. Rev. A}\ }\textbf {\bibinfo {volume} {73}},\ \bibinfo {pages} {013617} (\bibinfo {year} {2006})}\BibitemShut {NoStop}%
\bibitem [{\citenamefont {Hiller}, \citenamefont {Kottos},\ and\ \citenamefont {Geisel}(2006)}]{Hiller2006}%
  \BibitemOpen
  \bibfield  {author} {\bibinfo {author} {\bibfnamefont {M.}~\bibnamefont {Hiller}}, \bibinfo {author} {\bibfnamefont {T.}~\bibnamefont {Kottos}}, \ and\ \bibinfo {author} {\bibfnamefont {T.}~\bibnamefont {Geisel}},\ }\bibfield  {title} {\enquote {\bibinfo {title} {Complexity in parametric {B}ose-{H}ubbard {H}amiltonians and structural analysis of eigenstates},}\ }\href {\doibase 10.1103/PhysRevA.73.061604} {\bibfield  {journal} {\bibinfo  {journal} {Phys. Rev. A}\ }\textbf {\bibinfo {volume} {73}},\ \bibinfo {pages} {061604} (\bibinfo {year} {2006})}\BibitemShut {NoStop}%
\bibitem [{\citenamefont {Liu}\ \emph {et~al.}(2007)\citenamefont {Liu}, \citenamefont {Fu}, \citenamefont {Yang},\ and\ \citenamefont {Liu}}]{Liu2007}%
  \BibitemOpen
  \bibfield  {author} {\bibinfo {author} {\bibfnamefont {B.}~\bibnamefont {Liu}}, \bibinfo {author} {\bibfnamefont {L.-B.}\ \bibnamefont {Fu}}, \bibinfo {author} {\bibfnamefont {S.-P.}\ \bibnamefont {Yang}}, \ and\ \bibinfo {author} {\bibfnamefont {J.}~\bibnamefont {Liu}},\ }\bibfield  {title} {\enquote {\bibinfo {title} {{J}osephson oscillation and transition to self-trapping for {B}ose-{E}instein condensates in a triple-well trap},}\ }\href {\doibase 10.1103/PhysRevA.75.033601} {\bibfield  {journal} {\bibinfo  {journal} {Phys. Rev. A}\ }\textbf {\bibinfo {volume} {75}},\ \bibinfo {pages} {033601} (\bibinfo {year} {2007})}\BibitemShut {NoStop}%
\bibitem [{\citenamefont {Hiller}, \citenamefont {Kottos},\ and\ \citenamefont {Geisel}(2009)}]{Hiller2009}%
  \BibitemOpen
  \bibfield  {author} {\bibinfo {author} {\bibfnamefont {M.}~\bibnamefont {Hiller}}, \bibinfo {author} {\bibfnamefont {T.}~\bibnamefont {Kottos}}, \ and\ \bibinfo {author} {\bibfnamefont {T.}~\bibnamefont {Geisel}},\ }\bibfield  {title} {\enquote {\bibinfo {title} {Wave-packet dynamics in energy space of a chaotic trimeric {B}ose-{H}ubbard system},}\ }\href {\doibase 10.1103/PhysRevA.79.023621} {\bibfield  {journal} {\bibinfo  {journal} {Phys. Rev. A}\ }\textbf {\bibinfo {volume} {79}},\ \bibinfo {pages} {023621} (\bibinfo {year} {2009})}\BibitemShut {NoStop}%
\bibitem [{\citenamefont {Kollath}\ \emph {et~al.}(2010)\citenamefont {Kollath}, \citenamefont {Roux}, \citenamefont {Biroli},\ and\ \citenamefont {Läuchli}}]{Kollath2010}%
  \BibitemOpen
  \bibfield  {author} {\bibinfo {author} {\bibfnamefont {C.}~\bibnamefont {Kollath}}, \bibinfo {author} {\bibfnamefont {G.}~\bibnamefont {Roux}}, \bibinfo {author} {\bibfnamefont {G.}~\bibnamefont {Biroli}}, \ and\ \bibinfo {author} {\bibfnamefont {A.~M.}\ \bibnamefont {Läuchli}},\ }\bibfield  {title} {\enquote {\bibinfo {title} {Statistical properties of the spectrum of the extended {B}ose–{H}ubbard model},}\ }\href {\doibase 10.1088/1742-5468/2010/08/P08011} {\bibfield  {journal} {\bibinfo  {journal} {Journal of Statistical Mechanics: Theory and Experiment}\ }\textbf {\bibinfo {volume} {2010}},\ \bibinfo {pages} {P08011} (\bibinfo {year} {2010})}\BibitemShut {NoStop}%
\bibitem [{\citenamefont {Viscondi}\ and\ \citenamefont {Furuya}(2011)}]{Viscondi2011}%
  \BibitemOpen
  \bibfield  {author} {\bibinfo {author} {\bibfnamefont {T.~F.}\ \bibnamefont {Viscondi}}\ and\ \bibinfo {author} {\bibfnamefont {K.}~\bibnamefont {Furuya}},\ }\bibfield  {title} {\enquote {\bibinfo {title} {Dynamics of a {B}ose–{E}instein condensate in a symmetric triple-well trap},}\ }\href {\doibase 10.1088/1751-8113/44/17/175301} {\bibfield  {journal} {\bibinfo  {journal} {Journal of Physics A: Mathematical and Theoretical}\ }\textbf {\bibinfo {volume} {44}},\ \bibinfo {pages} {175301} (\bibinfo {year} {2011})}\BibitemShut {NoStop}%
\bibitem [{\citenamefont {Garcia-March}\ \emph {et~al.}(2018)\citenamefont {Garcia-March}, \citenamefont {van Frank}, \citenamefont {Bonneau}, \citenamefont {Schmiedmayer}, \citenamefont {Lewenstein},\ and\ \citenamefont {Santos}}]{March2018}%
  \BibitemOpen
  \bibfield  {author} {\bibinfo {author} {\bibfnamefont {M.~A.}\ \bibnamefont {Garcia-March}}, \bibinfo {author} {\bibfnamefont {S.}~\bibnamefont {van Frank}}, \bibinfo {author} {\bibfnamefont {M.}~\bibnamefont {Bonneau}}, \bibinfo {author} {\bibfnamefont {J.}~\bibnamefont {Schmiedmayer}}, \bibinfo {author} {\bibfnamefont {M.}~\bibnamefont {Lewenstein}}, \ and\ \bibinfo {author} {\bibfnamefont {L.~F.}\ \bibnamefont {Santos}},\ }\bibfield  {title} {\enquote {\bibinfo {title} {Relaxation, chaos, and thermalization in a three-mode model of a {B}ose{\textendash}{E}instein condensate},}\ }\href {\doibase 10.1088/1367-2630/aaed68} {\bibfield  {journal} {\bibinfo  {journal} {New J. Phys.}\ }\textbf {\bibinfo {volume} {20}},\ \bibinfo {pages} {113039} (\bibinfo {year} {2018})}\BibitemShut {NoStop}%
\bibitem [{\citenamefont {Bera}\ \emph {et~al.}(2019)\citenamefont {Bera}, \citenamefont {Roy}, \citenamefont {Gammal}, \citenamefont {Chakrabarti},\ and\ \citenamefont {Chatterjee}}]{Bera2019}%
  \BibitemOpen
  \bibfield  {author} {\bibinfo {author} {\bibfnamefont {S.}~\bibnamefont {Bera}}, \bibinfo {author} {\bibfnamefont {R.}~\bibnamefont {Roy}}, \bibinfo {author} {\bibfnamefont {A.}~\bibnamefont {Gammal}}, \bibinfo {author} {\bibfnamefont {B.}~\bibnamefont {Chakrabarti}}, \ and\ \bibinfo {author} {\bibfnamefont {B.}~\bibnamefont {Chatterjee}},\ }\bibfield  {title} {\enquote {\bibinfo {title} {Probing relaxation dynamics of a few strongly correlated bosons in a 1d triple well optical lattice},}\ }\href {\doibase 10.1088/1361-6455/ab2999} {\bibfield  {journal} {\bibinfo  {journal} {Journal of Physics B: Atomic, Molecular and Optical Physics}\ }\textbf {\bibinfo {volume} {52}},\ \bibinfo {pages} {215303} (\bibinfo {year} {2019})}\BibitemShut {NoStop}%
\bibitem [{\citenamefont {Rautenberg}\ and\ \citenamefont {G\"arttner}(2020)}]{Rautenberg2020}%
  \BibitemOpen
  \bibfield  {author} {\bibinfo {author} {\bibfnamefont {M.}~\bibnamefont {Rautenberg}}\ and\ \bibinfo {author} {\bibfnamefont {M.}~\bibnamefont {G\"arttner}},\ }\bibfield  {title} {\enquote {\bibinfo {title} {Classical and quantum chaos in a three-mode bosonic system},}\ }\href {\doibase 10.1103/PhysRevA.101.053604} {\bibfield  {journal} {\bibinfo  {journal} {Phys. Rev. A}\ }\textbf {\bibinfo {volume} {101}},\ \bibinfo {pages} {053604} (\bibinfo {year} {2020})}\BibitemShut {NoStop}%
\bibitem [{\citenamefont {Ray}, \citenamefont {Cohen},\ and\ \citenamefont {Vardi}(2020)}]{Ray2020}%
  \BibitemOpen
  \bibfield  {author} {\bibinfo {author} {\bibfnamefont {S.}~\bibnamefont {Ray}}, \bibinfo {author} {\bibfnamefont {D.}~\bibnamefont {Cohen}}, \ and\ \bibinfo {author} {\bibfnamefont {A.}~\bibnamefont {Vardi}},\ }\bibfield  {title} {\enquote {\bibinfo {title} {Chaos-induced breakdown of {B}ose-{H}ubbard modeling},}\ }\href {\doibase 10.1103/PhysRevA.101.013624} {\bibfield  {journal} {\bibinfo  {journal} {Phys. Rev. A}\ }\textbf {\bibinfo {volume} {101}},\ \bibinfo {pages} {013624} (\bibinfo {year} {2020})}\BibitemShut {NoStop}%
\bibitem [{\citenamefont {Nakerst}\ and\ \citenamefont {Haque}(2021)}]{Nakerst2021}%
  \BibitemOpen
  \bibfield  {author} {\bibinfo {author} {\bibfnamefont {G.}~\bibnamefont {Nakerst}}\ and\ \bibinfo {author} {\bibfnamefont {M.}~\bibnamefont {Haque}},\ }\bibfield  {title} {\enquote {\bibinfo {title} {Eigenstate thermalization scaling in approaching the classical limit},}\ }\href {\doibase 10.1103/PhysRevE.103.042109} {\bibfield  {journal} {\bibinfo  {journal} {Phys. Rev. E}\ }\textbf {\bibinfo {volume} {103}},\ \bibinfo {pages} {042109} (\bibinfo {year} {2021})}\BibitemShut {NoStop}%
\bibitem [{\citenamefont {Nakerst}\ and\ \citenamefont {Haque}(2023)}]{NH2022}%
  \BibitemOpen
  \bibfield  {author} {\bibinfo {author} {\bibfnamefont {G.}~\bibnamefont {Nakerst}}\ and\ \bibinfo {author} {\bibfnamefont {M.}~\bibnamefont {Haque}},\ }\bibfield  {title} {\enquote {\bibinfo {title} {Chaos in the three-site {B}ose-{H}ubbard model: Classical versus quantum},}\ }\href {\doibase 10.1103/PhysRevE.107.024210} {\bibfield  {journal} {\bibinfo  {journal} {Phys. Rev. E}\ }\textbf {\bibinfo {volume} {107}},\ \bibinfo {pages} {024210} (\bibinfo {year} {2023})}\BibitemShut {NoStop}%
\bibitem [{\citenamefont {Bhattacharyya}, \citenamefont {Ghosh},\ and\ \citenamefont {Nandi}(2023)}]{Bhattacharyya2023}%
  \BibitemOpen
  \bibfield  {author} {\bibinfo {author} {\bibfnamefont {A.}~\bibnamefont {Bhattacharyya}}, \bibinfo {author} {\bibfnamefont {D.}~\bibnamefont {Ghosh}}, \ and\ \bibinfo {author} {\bibfnamefont {P.}~\bibnamefont {Nandi}},\ }\bibfield  {title} {\enquote {\bibinfo {title} {Operator growth and krylov complexity in {B}ose-{H}ubbard model},}\ }\href {\doibase 10.1007/JHEP12(2023)112} {\bibfield  {journal} {\bibinfo  {journal} {J. High Energ. Phys.}\ }\textbf {\bibinfo {volume} {112}},\ \bibinfo {pages} {2023} (\bibinfo {year} {2023})}\BibitemShut {NoStop}%
\bibitem [{\citenamefont {Zhou}\ and\ \citenamefont {Chen}(2024)}]{Zhou24}%
  \BibitemOpen
  \bibfield  {author} {\bibinfo {author} {\bibfnamefont {B.}~\bibnamefont {Zhou}}\ and\ \bibinfo {author} {\bibfnamefont {S.}~\bibnamefont {Chen}},\ }\bibfield  {title} {\enquote {\bibinfo {title} {Spread complexity and dynamical transition in multimode bose-einstein condensates},}\ }\href {\doibase 10.1103/PhysRevB.110.064318} {\bibfield  {journal} {\bibinfo  {journal} {Phys. Rev. B}\ }\textbf {\bibinfo {volume} {110}},\ \bibinfo {pages} {064318} (\bibinfo {year} {2024})}\BibitemShut {NoStop}%
\bibitem [{\citenamefont {Robnik}(2019)}]{Robnik2019}%
  \BibitemOpen
  \bibfield  {author} {\bibinfo {author} {\bibfnamefont {M.}~\bibnamefont {Robnik}},\ }\enquote {\bibinfo {title} {Recent advances in quantum chaos of generic systems},}\ in\ \href {\doibase 10.1007/978-3-642-27737-5_730-1} {\emph {\bibinfo {booktitle} {Encyclopedia of Complexity and Systems Science}}},\ \bibinfo {editor} {edited by\ \bibinfo {editor} {\bibfnamefont {R.~A.}\ \bibnamefont {Meyers}}}\ (\bibinfo  {publisher} {Springer Berlin Heidelberg},\ \bibinfo {address} {Berlin, Heidelberg},\ \bibinfo {year} {2019})\ pp.\ \bibinfo {pages} {1--17}\BibitemShut {NoStop}%
\bibitem [{\citenamefont {Robnik}(2020)}]{Robnik2020}%
  \BibitemOpen
  \bibfield  {author} {\bibinfo {author} {\bibfnamefont {M.}~\bibnamefont {Robnik}},\ }\bibfield  {title} {\enquote {\bibinfo {title} {A brief introduction to stationary quantum chaos in generic systems},}\ }\href {\doibase https://doi.org/10.33581/1561-4085-2020-23-2-172-191} {\bibfield  {journal} {\bibinfo  {journal} {Nonlinear Phenomena in Complex Systems}\ }\textbf {\bibinfo {volume} {23}},\ \bibinfo {pages} {172--191} (\bibinfo {year} {2020})}\BibitemShut {NoStop}%
\bibitem [{Note1()}]{Note1}%
  \BibitemOpen
  \bibinfo {note} {Through Eq. \protect \eqref {eq:hclass1}, with \({\protect \mathcal {H}}_{\protect \text {cl}} = E_{\protect \mathrm {classic}}\), we determine suitable initial conditions for classical trajectories with energy \(E_{\protect \mathrm {classic}}\). A variant of Eq. \protect \eqref {eq:hclass1}, incorporating the coordinate \(\rho _2\) and Lagrange multipliers, is employed to find the classical critical configurations and their corresponding energies~\cite {Castro21}.}\BibitemShut {Stop}%
\bibitem [{\citenamefont {Santos}, \citenamefont {T\'avora},\ and\ \citenamefont {P\'erez-Bernal}(2016)}]{Santos2016}%
  \BibitemOpen
  \bibfield  {author} {\bibinfo {author} {\bibfnamefont {L.~F.}\ \bibnamefont {Santos}}, \bibinfo {author} {\bibfnamefont {M.}~\bibnamefont {T\'avora}}, \ and\ \bibinfo {author} {\bibfnamefont {F.}~\bibnamefont {P\'erez-Bernal}},\ }\bibfield  {title} {\enquote {\bibinfo {title} {{E}xcited-state quantum phase transitions in many-body systems with infinite-range interaction: {L}ocalization, dynamics, and bifurcation},}\ }\href {\doibase 10.1103/PhysRevA.94.012113} {\bibfield  {journal} {\bibinfo  {journal} {Phys. Rev. A}\ }\textbf {\bibinfo {volume} {94}},\ \bibinfo {pages} {012113} (\bibinfo {year} {2016})}\BibitemShut {NoStop}%
\bibitem [{\citenamefont {Mehta}(2004)}]{mehta2004random}%
  \BibitemOpen
  \bibfield  {author} {\bibinfo {author} {\bibfnamefont {M.~L.}\ \bibnamefont {Mehta}},\ }\href@noop {} {\emph {\bibinfo {title} {Random matrices}}}\ (\bibinfo  {publisher} {Elsevier},\ \bibinfo {year} {2004})\BibitemShut {NoStop}%
\bibitem [{\citenamefont {Bohigas}, \citenamefont {Giannoni},\ and\ \citenamefont {Schmit}(1984)}]{BGC1984}%
  \BibitemOpen
  \bibfield  {author} {\bibinfo {author} {\bibfnamefont {O.}~\bibnamefont {Bohigas}}, \bibinfo {author} {\bibfnamefont {M.~J.}\ \bibnamefont {Giannoni}}, \ and\ \bibinfo {author} {\bibfnamefont {C.}~\bibnamefont {Schmit}},\ }\bibfield  {title} {\enquote {\bibinfo {title} {Characterization of chaotic quantum spectra and universality of level fluctuation laws},}\ }\href {\doibase 10.1103/PhysRevLett.52.1} {\bibfield  {journal} {\bibinfo  {journal} {Phys. Rev. Lett.}\ }\textbf {\bibinfo {volume} {52}},\ \bibinfo {pages} {1--4} (\bibinfo {year} {1984})}\BibitemShut {NoStop}%
\end{thebibliography}

%

\end{document}